\documentclass[]{article}

\usepackage{multirow, amsmath, amsthm, graphicx, amssymb, amsfonts, latexsym, enumerate, amscd, xcolor, float, caption, bbold}
\usepackage{listings}
\usepackage{geometry}
\RequirePackage[numbers]{natbib}

\newtheorem{definition}{Definition}
\newtheorem{proposition}{Proposition}

\DeclareMathOperator{\sign}{sign}
\DeclareMathOperator{\diag}{diag} 
\DeclareMathOperator{\La}{LA} 
 
\DeclareMathOperator{\G}{G} 
\DeclareMathOperator{\IG}{IG} 
\DeclareMathOperator{\SGC}{SGC} 
\DeclareMathOperator{\E}{E} 
\DeclareMathOperator{\V}{V} 
\DeclareMathOperator{\Cov}{Cov} 

\usepackage[title]{appendix}

\usepackage{etoolbox}
\apptocmd{\appendices}{\apptocmd{\thesection}{:}{}{}}{}{}

\begin{document}
\title{Joint Posterior Inference for Latent Gaussian Models with R-INLA}

\author{Cristian Chiuchiolo (cristian.chiuchiolo@kaust.edu.sa )\\
AND \\
Janet van Niekerk (janet.vanNiekerk@kaust.edu.sa ) \\
AND \\
H\aa vard Rue (haavard.rue@kaust.edu.sa )\\
CEMSE Division\\
	King Abdullah University of Science and Technology\\
	Kingdom of Saudi Arabia}

\maketitle

\begin{abstract}
Efficient Bayesian inference remains a computational challenge in hierarchical models. Simulation-based approaches such as Markov Chain Monte Carlo methods are still popular but have a large computational cost. When dealing with the large class of Latent Gaussian Models, the INLA methodology embedded in the R-INLA software provides accurate Bayesian inference by computing deterministic mixture representation to approximate the joint posterior, from which marginals are computed. \\ The INLA approach has from the beginning been targeting to approximate univariate  posteriors. In this paper we lay out the development foundation of the tools for also providing joint approximations for subsets of the latent field. These approximations inherit Gaussian copula structure and additionally provide corrections for skewness. The same idea is carried forward also to sampling from the mixture representation, which we now can adjust for skewness. 
\end{abstract}

\section{Introduction}

Bayesian inference for latent Gaussian models (LGMs) have been made more attainable for high dimension and/or complex hierarchical models by the introduction of the Integrated Nested Laplace Approximation (INLA) method proposed by Rue et al (2009). Henceforth this methodology has been used extensively in various fields and modeling applications, see amongst others: health data applications in \cite{alvaro2013, yli2012, seppa2018}; spline models applied in medicine in \cite{bauer2016} or spatial/spatio-temporal models in \cite{beguin2012, gomezrubio2021, gomez2019, yyuan2017, peluso2020, meehan2019, pereira2017}; measurement error model in \cite{muff2015}; modelling applications in \cite{quiroz2015, ferkingstad2017, sorbye2019} and air data in \cite{dawkins2019}; a functional data analysis in \cite{yyue2019} and time trends for related populations in \cite{riebler2012est, riebler2012gender}; environmental data applications in \cite{huang2017} and \cite{illian2012} with some genetics in \cite{holand2013}; dynamic and stochastic volatility models respectively in \cite{rruiz2012, martino2010est}; joint models and survival applications in \cite{martino2010approx, vanniekerk2019new, rustand2021bayesian}. It is clear that the advancement made by INLA in the field of deterministic (not sampling-based) Bayesian inference, holds significant impact in science as a whole. For this reason, the developers continue to improve current implementations and develop additional tools for the data analysts' toolbox. \\ This paper will discuss how the marginal inference can be further extended to a joint one by exploiting a mixture representation in terms of the full conditional latent density of a Latent Gaussian Model. For readers not accustomed to INLA and all its features, we strongly suggest to take a look at some preliminaries given in Section \ref{sec:0} for a brief background before moving on. \\ First, it is essential to recall that the joint posterior density of a Latent Gaussian model with data $\boldsymbol{y}$ with likelihood hyperparameters $\boldsymbol{\theta_y}$ and latent field $\boldsymbol{x}$ with centered Gaussian prior with precision matrix $\boldsymbol{Q(\theta_x)}$, can be formalized into the following hierarchical structure

\begin{align}
\pi(\boldsymbol{x},\boldsymbol{\theta} \vert \boldsymbol{y}) &\propto \pi(\boldsymbol{\theta}) \pi(\boldsymbol{x} \vert \boldsymbol{\theta_x}) \prod_{i \in \mathcal{I}} \pi(y_i \vert x_i,\boldsymbol{\theta_y}) \nonumber \\
&\propto \pi(\boldsymbol{\theta}) \vert \boldsymbol{Q(\theta_x)} \vert^{\frac{1}{2}} \exp \Bigl ( -\frac{1}{2} \boldsymbol{x}^T \boldsymbol{Q(\theta_x)} \boldsymbol{x} + \sum_{i=1}^n \log [\pi(y_i|x_i, \boldsymbol{\theta_y})]\Bigr )
\label{joint:eq}
\end{align}

\noindent
with prior components on the hyperparameter set $\boldsymbol{\theta} = \{\boldsymbol{\theta_x},\boldsymbol{\theta_y}\}$, and likelihood contribution from a set of observed response $\mathcal{I}$ with dimensions $\vert \boldsymbol{x} \vert = N$ and $\vert  \mathcal{I} \vert = n$. The hyperparameter set is generally assumed to be small (let's say less than $10-15$) while the latent field contains both parameter and linear predictor terms that require inferential analysis. \\ Working on LGMs using INLA means that we are mainly interested in the  posterior marginals of their latent field $\boldsymbol{x}$ which can be obtained from the joint density in~\eqref{joint:eq} as follows

\begin{align}
    \pi(\boldsymbol{x} \vert \boldsymbol{y}) &=  \int \pi(\boldsymbol{x}, \boldsymbol{\theta} \vert \boldsymbol{y}) \, d \boldsymbol{\theta} \nonumber \\
    &= \int \pi(\boldsymbol{x} \vert \boldsymbol{\theta}, \boldsymbol{y}) \pi(\boldsymbol{\theta} \vert \boldsymbol{y}) \, d \boldsymbol{\theta}
\label{marg_lat}
\end{align}

\noindent
Approximations for these marginals are achieved through the use of Laplace approximations on both density components defined at the right side of \eqref{marg_lat} and then integrating out the uncertainty coming from the hyperparameter set $\boldsymbol{\theta}$: this is acknowledged as the Nested Laplace scheme of INLA. \\ In many cases where the likelihood contribution of~\eqref{joint:eq} is Gaussian or relatively close to that shape, we can make use of the most accurate and fast Gaussian approximation onto the density $\pi(\boldsymbol{x} \vert \boldsymbol{\theta}, \boldsymbol{y})$. This joint Gaussian approximation, which we denote as $\tilde{\pi}_{\G}(\boldsymbol{x} \vert \boldsymbol{\theta}, \boldsymbol{y})$, represents the central key point to this entire work. \\ \\ As the name suggests, the Gaussian approximation relates to a multivariate Normal density structure with mean and correlation structure inherited by the prior information assumed in~\eqref{joint:eq} from both the Gaussian latent field and the likelihood density of observed data $\boldsymbol{y}$. \\ However this joint approximation is bounded by its intrinsic construction which can lack accuracy in more extreme settings where Gaussian assumptions may not hold. \\ Such issues can be solved by introducing a new joint approximation which contains corrections for marginal location and skewness in the case of substantial deviation from Gaussian behaviour. \\ While taking into account the Gaussian approximation on $\boldsymbol{x}$, we define a set of well defined transformations $\boldsymbol{\tilde{x}}=\boldsymbol{h(x)}=( h_1(x_1), \dots, h_N(x_N) )$ for the entire latent field and envelope the entire joint object into a Gaussian copula structure (see \cite{nelsen_1999_an} for more general details). The resulting joint density can be formalized as

\begin{align}
    \tilde{\pi}_{\SGC}(\boldsymbol{\tilde{x}} \vert \boldsymbol{\theta}, \boldsymbol{y}) &\propto \tilde{\pi}_{\G}(\boldsymbol{h}^{-1}(\boldsymbol{\tilde{x}}) \vert \boldsymbol{\theta}, \boldsymbol{y}) \vert \boldsymbol{J_{\tilde{x}}} \vert \nonumber \\
    &\propto \vert \boldsymbol{J_{\tilde{x}}} \vert \exp(-\frac{1}{2}[\boldsymbol{h}^{-1}(\boldsymbol{\tilde{x}})-\boldsymbol{\mu}^*(\boldsymbol{\theta})]^T \boldsymbol{Q}^*(\boldsymbol{\theta})[\boldsymbol{h}^{-1}(\boldsymbol{\tilde{x}})-\boldsymbol{\mu}^*(\boldsymbol{\theta})] 
    \label{gau_tran}
\end{align}

\noindent
where $(\boldsymbol{\mu}^*(\boldsymbol{\theta}), \boldsymbol{Q^*(\theta)})$ are the posterior summaries from the Gaussian approximation while $ \boldsymbol{J_{\tilde{x}}}$ is the Jacobian belonging to the transformation $\boldsymbol{h(\cdot)}$. Here we recall that $\boldsymbol{\mu}^*(\boldsymbol{\theta})$ comes from matching the mode of $\pi(\boldsymbol{x} \vert \boldsymbol{\theta}, \boldsymbol{y})$ and $\boldsymbol{Q}^*(\boldsymbol{\theta})=\boldsymbol{Q}(\boldsymbol{\theta})+\diag [\boldsymbol{c}(\boldsymbol{\theta})]$ where $\boldsymbol{Q}(\boldsymbol{\theta})$ comes from the latent field assumption while $\boldsymbol{c}(\boldsymbol{\theta})$ contains the negative second derivative of the log likelihood. \\ The density in~\eqref{gau_tran} strongly depends on the choice of the set of functions $\boldsymbol{h(\cdot)}$ which mostly aim to define skewed marginal transformations. Such joint construction allows to properly model density assumptions that deviate from a Gaussian behaviour while retaining the dependency structure amongst the latent field terms through the aid of a Gaussian copula. \\ The multivariate object we get is then acknowledged as a new class of joint approximation densities named \emph{Skew Gaussian copula} (SGC). \\ To point out that a smart choice of the set of marginal transformations $\boldsymbol{h(\cdot)}$ can grant more accurate corrections for location and skewness therefore dealing with more extreme cases. \\ Such settings can be count data with unbalanced low counts or low number of successes in a binomial experiment where extreme marginal skewness behaviour are likely to arise and can be hard to detect with the Gaussian approximation as it is. \\ \\ As a brief example, let us consider Poisson observations $y_1, \dots, y_n$ with means $\lambda_1, \dots, \lambda_n$, a single covariate $\boldsymbol{\xi}=( \xi_1, \dots, \xi_n )$ and linear predictors $\eta_i = \log \lambda_i=\beta_0+\beta_1 \xi_i$, $\forall i$ with $(\beta_0, \beta_1)$ being the coefficients for the intercept and covariate respectively. According to INLA methodology, the latent field vector would be $\boldsymbol{x} = (\eta_1, \dots, \eta_n, \beta_0, \beta_1)$ with $\boldsymbol{x} \sim N(\boldsymbol{0}, \boldsymbol{Q})$ where $\boldsymbol{Q}$ is a symmetric precision matrix with dimension $(n+2)\times (n+2)$. \\ In this case we would have $\pi(\boldsymbol{\theta} \vert \boldsymbol{y}) \propto 1$ since no hyperparameter is assumed in the model structure and 

\begin{align}
    \pi(\boldsymbol{x} \vert \boldsymbol{y}) &\propto \pi(\boldsymbol{y} \vert \boldsymbol{x}) \pi(\boldsymbol{x}) \nonumber \\
    & \propto \exp \Bigl ( -\frac{1}{2} \boldsymbol{x}^T \boldsymbol{Qx}+\sum_{i=1}^n [y_i\eta_i-\exp(\eta_i)]\Bigr ) 
    \label{poi_ex_joint}
\end{align}
which represents the full joint posterior density of the model, as it is. Accordingly, the precision matrix $\boldsymbol{Q}$ would be

\begin{equation}
    \boldsymbol{Q} = \begin{pmatrix}
\tau_{\epsilon} \boldsymbol{I} & \tau_{\epsilon} \boldsymbol{I \xi} & \tau_{\epsilon} \boldsymbol{I} \boldsymbol{\mathbb{1}}\\ 
 & \tau_{\beta_0}+\tau_{\epsilon} \boldsymbol{\mathbb{1}}^T\boldsymbol{\mathbb{1}} & \tau_{\epsilon} \boldsymbol{\xi}^T  \boldsymbol{\mathbb{1}}\\
 &  & \tau_{\beta_1} +\tau_{\epsilon} \boldsymbol{\xi}^T\boldsymbol{\xi}\\
\end{pmatrix}
\label{Qstru}
\end{equation}
where $\boldsymbol{I}$ is a $n \times n$  identity matrix, $\mathbb{1}=(1,1,\dots,1)^T$ a $(n+2)$-dimensional unit vector and $(\tau_{\beta_0}, \tau_{\beta_1})$ are the precisions of the fixed parameters $(\beta_0, \beta_1)$. The precision $\tau_{\epsilon}$ appearing in structure~\eqref{Qstru} relates to a tiny Gaussian noise $\epsilon$ added to each linear predictor $\eta_i$ to avoid singularity issues for $\boldsymbol{Q}^{-1}$ (see \cite{rue2017computing} for a more general representation). \\ The respective Gaussian approximation of~\eqref{poi_ex_joint} would be $\tilde{\pi}_{\G}(\boldsymbol{x} \vert \boldsymbol{y})$ with density $N(\boldsymbol{x}^*, \boldsymbol{Q}^*)$ where $\boldsymbol{x}^*=(x^*_1,\dots,x^*_n,x^*_{n+1}, x^*_{n+2})$ is the mean summary resulting from matching the modal configuration of $\pi(\boldsymbol{x} \vert \boldsymbol{y})$ while $\boldsymbol{Q}^*=\boldsymbol{Q}+\diag (\boldsymbol{c})$ with

\begin{equation}
    \boldsymbol{c} = \begin{pmatrix}
- \frac{\partial^2}{\partial^2 \eta_1} [\sum_{i=1}^n y_i\eta_i-\exp(\eta_i)] \vert_{\eta_1 = x^*_1} \\
\vdots \\
- \frac{\partial^2}{\partial^2 \eta_n} [\sum_{i=1}^n y_i\eta_i-\exp(\eta_i)] \vert_{\eta_n = x^*_n} \\
- \frac{\partial^2}{\partial^2 \beta_0} [\sum_{i=1}^n y_i(\beta_0+\beta_1 \xi_i)-\exp(\beta_0+\beta_1 \xi_i)] \vert_{\beta_0 = x^*_{n+1}} \\
- \frac{\partial^2}{\partial^2 \beta_1} [\sum_{i=1}^n y_i(\beta_0+\beta_1 \xi_i)-\exp(\beta_0+\beta_1 \xi_i)] \vert_{\beta_1 = x^*_{n+2}} \\
\end{pmatrix}
\end{equation}
which contains all the negative second derivatives of the log likelihood in~\eqref{poi_ex_joint} evaluated at the modal points within $\boldsymbol{x}^*$ with respect to each latent parameter $x_i$. \\ Since Poisson data may be inaccurately described by a Gaussian approximation, we may exploit  the general class of approximation densities in~\eqref{gau_tran} by considering a transformation $\boldsymbol{h(\cdot)}$ which can add more skewness adjustments to properly recover the true underlying model density. \\ By doing so, we would still be able to retain the dependency structure encoded in $\boldsymbol{Q}$ unscathed due to the use of the Gaussian copula while getting benefit of new marginal non Gaussian assumptions encoded in $\boldsymbol{h(\cdot)}$. \\ \\ Although the marginal posterior analysis is largely achievable with INLA by using Gaussian or Laplace approximations onto the posterior densities in ~\eqref{marg_lat}, in some other cases we may a different approach. When the hyperparameter information is strongly correlated to one or more parameters or is intrinsically part of the underlying model structure, then we might need to turn to a joint posterior inference. \\ As we have seen, the Gaussian approximation can satisfy most of the joint inference purposes but still lacks proper adjustments when the Gaussian assumption does not properly hold. \\ This is why in Section \ref{sec:1} we first spend some effort to properly define a new class of joint approximations which enlarges the set of approximations that INLA can use based on different choices of the transformation $\boldsymbol{h(\cdot)}$. Since these transformations allow to encode skewed behaviour into the Gaussian Copula structure, we name this class as \emph{Skew Gaussian Copula}. \\ Similarly, in Section \ref{sec:2}, we construct an alternative strategy that borrows the correlation structure and the summaries of the approximation in~\eqref{gau_tran} to achieve joint analytical approximations for a joint subset of the latent parameters terms. Such analytical approximation follows the same features of the Skew Gaussian Copula class by using Skew Normal densities for approximating marginals and can target joint densities $\pi(\boldsymbol{x}_S \vert \boldsymbol{y})$, with $S$ being a set of indexes of the latent field components, and additive linear combination densities of the form $\pi(\boldsymbol{Ax} \vert \boldsymbol{y})$ where $\boldsymbol{A}$ is a $M \times N$ matrix of indexes with $M$ being the number of linear combinations. \\ With this new tool, we are able to get fast and accurate results for both above-mentioned cases. \\ When such assumptions do not hold or the size of the model is too large, we can exploit the same relation in~\eqref{marg_lat} by not integrating out $\boldsymbol{\theta}$ and achieve a joint approximation for $\pi(\boldsymbol{x}, \boldsymbol{\theta} \vert \boldsymbol{y})$. Since this joint density is unknown and a deterministic approximation is unreachable at an overall level, we rely on a sampling based approach in terms of the same class of approximations $\tilde{\pi}_{\SGC}(\boldsymbol{\tilde{x}} \vert \boldsymbol{\theta}, \boldsymbol{y})$ as seen in~\eqref{gau_tran}. \\ In Section \ref{sec:3}, we lay down a formal derivation of the mixture structure which allows the joint posterior approximation to inherit all the features from this class. Proper corrected results are reached by drawing samples from this improved approximation. \\ Then, in Section \ref{sec:4} we present numerical illustrations to underline the intrinsic differences amongst these new tools which increase the variety of the approximation strategies. The results are accomplished by both using INLA, main software for the presented strategies of this work, and JAGS (\cite{plummer_2003_dsc}) regarding the MCMC sampling counterpart for the sake of comparison.

\section{Preliminaries on INLA}
\label{sec:0}

Here we provide a brief useful introduction of the INLA approach for tackling Bayesian inference. \\ We briefly introduce latent Gaussian models (LGM) and some details of the INLA methodology, sufficient for the work presented here. Again, we believe that the reader familiar with INLA would be comfortable to continue to the next section.\\
Latent Gaussian models are statistical models that relate the general (continuous,discrete or joint) response $\boldsymbol{y}$ to an additive linear predictor $\boldsymbol{\eta}$ while assuming a Gaussian Markov Random Field (GMRF, \cite{rue2005gmrf}) for the latent field, $\boldsymbol{x}$, of the model. This formulation allows great computational advantages for any dimension of the model with stable accuracy in the results. Some models in this class are GLM, GAMM, time, space-time, spline, survival and joint models. The linear predictor term for each observation is then

\begin{equation}
 \eta_i = \gamma_0 + \sum_{j=1}^{n_J} \gamma_j z_{ij} + \sum_{k=1}^{n_K}f_k(u_{ik}) +\epsilon_i \quad \text{for} \quad i=1,\dots, n
\label{pred_eq}
\end{equation}

\noindent
where each $\eta_i$ is related to one response $y_i$ only through a well-specified link function $g(\mu_i)=\eta_i$ for the mean component $\mu_i$. From ~\eqref{pred_eq} we see the possible structured terms from covariates that can be accommodated in the LGM framework: $\gamma_0$ refers to the overall intercept of the model, the $\{\gamma_j \}$s describe the fixed and random coefficients assigned to each covariate $z_{ij}$ while $\{f_k(u_{ik})\}$ represent unknown well-defined functions associated with the covariates $\{ u_{ik} \}$. The terms $n_J$ and $n_K$ denote the dimension of the covariate space (fixed and random effects, respectively) while $\epsilon_i$ is assumed to follow a Gaussian distribution. Since all linear predictor terms in~\eqref{pred_eq} are included in the latent field $\boldsymbol{x}$, its overall dimension would be equal to $N = n+n_P$ where $n_P=n_J+n_K+1$ denotes the number of fixed and random latent parameters of the model.\\ \\
A multivariate Gaussian prior, $\pi(\boldsymbol{x}|\boldsymbol{\theta})$ with a sparse precision matrix is assumed for the latent field.
Often, the likelihood and the $\{f_k(u_{ik})\}$ terms can have additional hyperparameters, we denote the entire hyperparameter set  by $\boldsymbol{\theta}=(\boldsymbol{\theta_x},\boldsymbol{\theta_y})$, where $\boldsymbol{\theta_x}$ is the hyperparameters for the latent field and $\boldsymbol{\theta_y}$ is the hyperparameters for the likelihood. The prior for $\boldsymbol{\theta}, \pi(\boldsymbol{\theta})$ does not need to be Gaussian, and various options are available in \textbf{R-INLA}. \\ \\  INLA can compute accurate approximations for the univariate posterior marginals of the unknown parameters for the joint model density in~\eqref{joint:eq} and these results typically represent our main interest in a more applied inferential context. First we derive an approximation for all posterior marginals of the hyperparameters $\{ \pi(\theta_j \vert \boldsymbol{y}), j=1, \dots, \vert \boldsymbol{\theta} \vert \} $ by employing a Laplace approximation on its joint parent as follows

\begin{equation}
    \tilde{\pi}_{\La}(\boldsymbol{\theta} \vert \boldsymbol{y}) \propto \frac{\pi(\boldsymbol{x}^*, \boldsymbol{\theta} \vert \boldsymbol{y})}{\tilde{\pi}_{\G}(\boldsymbol{x}^* \vert \boldsymbol{\theta},\boldsymbol{y})} \Bigg \vert_{\boldsymbol{x}^*=\boldsymbol{\mu(\theta)}}
\label{joint:hyper}
\end{equation}

\noindent
where $\tilde{\pi}_{\G}(\boldsymbol{x}^* \vert \boldsymbol{\theta},\boldsymbol{y})$ is the Gaussian approximation obtained by matching the mode and curvature at the mode of the full joint density $\pi(\boldsymbol{x} \vert \boldsymbol{\theta}, \boldsymbol{y})$ after an iterative process. Then the entire Laplace approximation for~\eqref{joint:hyper} is achieved by evaluating the ratio at the denominator mean $\boldsymbol{\mu(\theta)}$ so that the result is still a function of $\boldsymbol{\theta}$. \\ This initial step requires a smart and balanced exploration of the density in~\eqref{joint:hyper} which ends up in the computation of a certain number of configuration points $\{ \boldsymbol{\theta_k}, k=1,\dots, K \}$ that are then used to get the corresponding marginals by interpolation (\cite{martins2013new}). The last step focus on approximating the remaining densities $\{ \pi(x_i \vert \boldsymbol{y}), i = 1, \dots, N \}$ by using the pre-computed points $\boldsymbol{\theta}_k$ 

\begin{equation}
\tilde{\pi}_{\La}(x_i \vert \boldsymbol{y}) \approx \sum_{k=1}^K \tilde{\pi}_{\La}(x_i \vert \boldsymbol{y}, \boldsymbol{\theta}_k) \tilde{\pi}_{\La}(\boldsymbol{\theta}_k \vert \boldsymbol{y}) \Delta_k
\label{marg_inla}
\end{equation}

\noindent
where the densities $\tilde{\pi}_{\La}(x_i \vert \boldsymbol{y}, \boldsymbol{\theta}_k)$ can be computed with three different strategies: \emph{Gaussian Approximation}, \emph{Laplace Approximation} or \emph{Simplified Laplace Approximation}. \\ The Gaussian approximation is always the best option in terms of speed since it exploits the pre-computed joint density at the denominator in~\eqref{joint:hyper} but can have flaws in location and skewness adjustments as stated in \cite{martino2007}. \\ The Laplace approximation is more computationally intensive but ensures more accuracy while the Simplified Laplace strategy represents a faster version of the Laplace at the cost of some accuracy which is negligible in most cases. \\ By default INLA exploits skew normal densities to get the full conditional approximations $\pi_{\La}(x_i \vert \boldsymbol{y}, \boldsymbol{\theta}_k) $ in~\eqref{marg_inla} based on a Taylor expansion up to third order on the Laplace approximation strategy (see \cite{rue2009inla} for further details).  

\section{Skew Gaussian Copula} 
\label{sec:1}

When trying to approximate posterior marginal in INLA, the first kind of approximation we encounter is the Gaussian approximation as it appears in the entire Laplace approximation scheme: first for the hyperparameter marginals in~\eqref{joint:hyper} and then as an option for the full conditional latent densities in~\eqref{marg_inla}. \\ Based on the work of \cite{ferkingstad2015latent} we know that the Gaussian approximation may turn out to be relatively inaccurate when the likelihood contribution comes from non Gaussian assumption, such as Binomial or Poisson, in the context of GLMMs. The solution is to adjust the approximation, pointwise, by using a Gaussian copula combined with a well defined transformation $\boldsymbol{h(\cdot)}$ on $\boldsymbol{x}$ that borrows strength from the more accurate Laplace approximations computed by INLA. To fully understand the gain of this whole idea, we must tackle its methodology point by point.

\subsection{Methodology}

Let us recall how INLA encodes concepts like Gaussian approximation and constraints into the latent field component. By definition, the first INLA step is summarised by the Laplace approximation in~\eqref{joint:hyper} which requires to compute a Gaussian approximation of the denominator to actually approximate the full conditional $\pi(\boldsymbol{x} \vert \boldsymbol{\theta}, \boldsymbol{y})$ whose density is

\begin{equation}
    \pi(\boldsymbol{x} \vert \boldsymbol{\theta}, \boldsymbol{y}) \propto \exp \Bigl ( -\frac{1}{2} \boldsymbol{x}^T \boldsymbol{Q(\boldsymbol{\theta})x}+ \sum_{i=1}^n \log \pi(y_i \vert x_i, \boldsymbol{\theta}) \Bigr )
    \label{full_cond}
\end{equation}

\noindent
where $\pi(y_i \vert x_i, \boldsymbol{\theta})$ refers to each likelihood point assumption while $\boldsymbol{x} \sim N(\boldsymbol{0}, \boldsymbol{Q^{-1}(\theta)})$. \\ A quite simple but powerful trick to approximate such densities is to collect and evaluate the available information such that the resulting kernel resembles a Normal distribution. This approach leads to the Gaussian approximation density on~\eqref{full_cond}

\begin{equation}
    \tilde{\pi}_{\G}(\boldsymbol{x} \vert \boldsymbol{y}, \boldsymbol{\theta}) = (2\pi)^{-\frac{N}{2}} \vert \boldsymbol{Q}^*(\boldsymbol{\theta}) \vert^{\frac{1}{2}} \exp \Bigl ( -\frac{1}{2} (\boldsymbol{x}-\boldsymbol{\mu(\theta)})^T \boldsymbol{Q}^*(\boldsymbol{\theta}) (\boldsymbol{x}-\boldsymbol{\mu(\theta)})  \Bigr )
    \label{gau_dens}
\end{equation}
which exactly corresponds to a multivariate normal density with mean $(\boldsymbol{\mu(\theta)})$ and precision matrix $\boldsymbol{Q}^*(\boldsymbol{\theta})$ with $\boldsymbol{Q}^*(\boldsymbol{\theta})= \boldsymbol{Q}(\boldsymbol{\theta})+\text{diag}(\boldsymbol{c}(\boldsymbol{\theta}))$ where $\boldsymbol{c}(\boldsymbol{\theta})$ contains the negative second derivatives of the log density evaluated at the mode of~\eqref{full_cond}. \\ These summaries are attained through an iterative Newton Raphson process (see \cite{rue2009inla}). If there are linear constraints of the form $\boldsymbol{Cx}=\boldsymbol{e}$ with $\boldsymbol{C}$ and $\boldsymbol{e}$ being a $N \times k$ matrix with rank $k$ and a real vector respectively, then the mean of the approximation is replaced within the iterative process by using the expected value onto a sample from a constrained GMRF (for a detailed explanation of \emph{Gaussian Markov Random Fields}, see \cite{rue2005gmrf}) which is given  by the following 

\begin{equation}
    \boldsymbol{x}^c= \boldsymbol{x}-\boldsymbol{Q^{-1}C^T}(\boldsymbol{CQ^{-1}C^T})^{-1}(\boldsymbol{Cx-e})
\end{equation}
where $\boldsymbol{x}$ is the usual notation for an unconstrained GMRF sample. \\ These concepts need to be carefully considered when trying to construct a joint posterior approximation to $\pi(\boldsymbol{x} \vert \boldsymbol{\theta}, \boldsymbol{y})$ since we must account for any changes in the overall precision structure and possible constraints adjustments if there are any. \\ \\ 
A blind use of the Gaussian approximation, as it is, may lead to inaccuracies in some cases, as extensively pointed out by the work of \cite{ferkingstad2015latent}. The real issue comes out when the likelihood contribution is skewed or far from being Gaussian as it may appear in some GLMM settings with Binomial or Poisson modelling assumptions. This is where the authors lay down the foundation for a new class of approximations which aims to encode more location and skewness adjustments by borrowing strength from the existing and widely used Laplace approximations. \\ \\ Through the Gaussian approximation definition in~\eqref{gau_dens}, we construct a new class of joint approximation densities by defining a new latent random field $\boldsymbol{\tilde{x}}$ and a set of marginal transformations $\boldsymbol{\tilde{x}} = \boldsymbol{h(x)}$ applied on the original latent field. \\ By combining these transformations and a Gaussian copula we are able to achieve a joint distribution object with with the benefit of the more accurate Laplace approximated marginals used in~\eqref{marg_inla}. This improved marginal information can be easily encoded in the resulting joint density while retaining the original dependency structure in the precision matrix $\boldsymbol{Q}(\boldsymbol{\theta})$ of the Gaussian approximation. \\ Such class of approximations is properly defined when a choice of $\boldsymbol{h(\cdot)}$ is made within the Gaussian copula structure. \\ \\ Since we seek a new joint density approximation applied to the new random latent vector denoted as $\boldsymbol{\tilde{x}} = ( h_1(x_1), \dots, h_N(x_N) )$ through the function vector $\boldsymbol{h(\cdot)}$, we may better set some notation first. \\ Let us consider $\boldsymbol{\tilde{x}} \sim \boldsymbol{F}$ where $\boldsymbol{F}=( F_1, \dots,F_i, \dots, F_N )$ is the vector of cumulative distribution functions coming from Skew Normal densities so that we can encode skewness into our new approximation. Such choice is coherent to how INLA approximates marginal densities through its Simplified Laplace strategy.  \\ Accordingly, $\boldsymbol{\tilde{z}} \sim \boldsymbol{\tilde{F}}$ with $\boldsymbol{\tilde{z}}= \frac{\boldsymbol{\tilde{x}}-\boldsymbol{\tilde{\mu}(\theta)}}{\boldsymbol{\tilde{\sigma}(\theta)}}$ so that $\boldsymbol{\tilde{F}}$ is the vector of standard cumulative distribution functions of a Skew Normal density with improved mean $\boldsymbol{\tilde{\mu}(\theta)}$ and standard deviation $\boldsymbol{\tilde{\sigma}(\theta)}$. \\ Thus, we can now apply the Gaussian copula and achieve explicit expressions for the transformations $\boldsymbol{h(\cdot)}$, towards the goal of creating a new joint approximation density to~\eqref{gau_dens}:

\begin{enumerate}
\item Let $\boldsymbol{z}=\frac{\boldsymbol{x}-\boldsymbol{\mu(\theta)}}{\boldsymbol{\sigma(\theta)}}$ be the original standardized latent term with respect to its Gaussian approximation $\tilde{\pi}_{\G}(\boldsymbol{x} \vert \boldsymbol{\theta},\boldsymbol{y})$ with respective vectorial posterior summaries $(\boldsymbol{\mu(\theta)}, \boldsymbol{\sigma(\theta)})$. 
\item Let $\boldsymbol{\tilde{z}}= \boldsymbol{h(z)}=\boldsymbol{\tilde{F}}^{-1}(\boldsymbol{\Phi(z)})$ with $\boldsymbol{\Phi}=(\Phi_1, \dots, \Phi_i, \dots, \Phi_N)$ being the vector of standard cumulative Gaussian distribution functions where $ \boldsymbol{\Phi}_i  = \Phi$, $\forall i$ by definition
\item Thus we observe that $\Phi(z_i) \sim U(0,1)$, $\forall i$ by using the PIT theorem while its inverse application leads to $\boldsymbol{\tilde{z}} \sim \boldsymbol{\tilde{F}}$ meaning that 
\begin{align}
\boldsymbol{\tilde{x}} &= \boldsymbol{\tilde{\sigma}(\theta)h(z)}+\boldsymbol{\tilde{\mu}(\theta)} \nonumber \\
&= \boldsymbol{\tilde{\sigma}(\theta)}\boldsymbol{\tilde{F}}^{-1}(\boldsymbol{\Phi(z)})+\boldsymbol{\tilde{\mu}(\theta)} 
\label{meancorr1}
\end{align}
where $\boldsymbol{\tilde{x}} \sim \boldsymbol{F}$ as per prior assumption. 
\end{enumerate}

\noindent
The explicit definition of the transformation $\boldsymbol{h(\cdot)}$ in point 2 above opens a series of features for the new class of joint approximation, named \emph{Skew Gaussian Copula} (SGC), due to the main use of the set of Skew Normal marginals $\boldsymbol{\tilde{F}}$ borrowed from the Simplified Laplace strategy. \\ In this way, the Skew Gaussian Copula can properly adjust its marginal properties based on the required accuracy by introducing location and skewness adjustments. \\ The explicit expression of the transformation $\boldsymbol{h(\cdot)}$ as

\begin{equation}
    \boldsymbol{h(z)} = \boldsymbol{\tilde{F}}^{-1} \Bigl [ \boldsymbol{\Phi} \Bigl ( \frac{\boldsymbol{x}-\boldsymbol{\mu}(\boldsymbol{\theta})}{\boldsymbol{\sigma}(\boldsymbol{\theta})}\Bigr )\Bigr]
    \label{hrel}
\end{equation}

\noindent
and its inverse transformation by change of variable theorem

\begin{equation}
\boldsymbol{z}=\boldsymbol{h}^{-1} ( \boldsymbol{\tilde{z}})= \boldsymbol{\Phi}^{-1} \Bigl [ \boldsymbol{\tilde{F}} \Bigl ( \frac{\boldsymbol{\tilde{x}}-\boldsymbol{\tilde{\mu}(\theta)}}{\boldsymbol{\tilde{\sigma}(\theta)}}\Bigr ) \Bigr ]
    \label{hinv}
\end{equation}

\noindent
lead to properly define the joint density of the Skew Gaussian Copula onto the new latent field $\tilde{\boldsymbol{x}}$ as follows

\begin{align}
    \tilde{\pi}_{\SGC}(\boldsymbol{\tilde{x}} \vert \boldsymbol{\theta}, \boldsymbol{y}) &= \tilde{\pi}_{\G}(\boldsymbol{g(\tilde{x})} \vert \boldsymbol{\theta}, \boldsymbol{y}) \vert \boldsymbol{J_x}  \vert \nonumber \\
    &= (2\pi)^{-\frac{N}{2}} \vert \boldsymbol{Q^*(\theta)} \vert^{\frac{1}{2}} \exp \Bigl [ -\frac{1}{2} ( \boldsymbol{g(\tilde{x})}-\boldsymbol{\mu(\theta)})^T \boldsymbol{Q^*(\theta)} (\boldsymbol{g(\tilde{x})}-\boldsymbol{\mu(\theta)}) \Bigr ] \vert \boldsymbol{J_x} \vert
    \label{fullden1}
\end{align}

\noindent
with $\boldsymbol{g(\tilde{x})}=\boldsymbol{h}^{-1} \Bigl ( \frac{\boldsymbol{\tilde{x}}-\boldsymbol{\tilde{\mu}(\theta)}}{\boldsymbol{\tilde{\sigma}(\theta)}} \Bigr )\boldsymbol{\sigma(\theta)}+\boldsymbol{\mu(\theta)}$ and $\vert \boldsymbol{J_x} \vert$ being the Jacobian determinant of the applied vectorial transformation  

\begin{equation}
    \boldsymbol{J_x} = \Bigl [ \frac{\partial \boldsymbol{x}}{\partial \tilde{x}_1}, \dots, \frac{\partial \boldsymbol{x}}{\partial \tilde{x}_N}\Bigr ] = \begin{pmatrix}
\frac{\partial x_1}{\partial \tilde{x}_1} & 0 & \dots & 0\\ 
0 & \frac{\partial x_2}{\partial \tilde{x}_2} & 0 & \vdots \\
\vdots & 0 & \ddots & 0 \\
0 & \dots & 0 & \frac{\partial x_N}{\partial \tilde{x}_N} 
\end{pmatrix}
\end{equation}
with $\vert \boldsymbol{J_x} \vert= \Bigl \vert \prod_i \frac{\partial x_i}{\partial \tilde{x}_i} \Bigr \vert $ where $\frac{\partial x_i}{\partial \tilde{x}_i} \ge 0$, $\forall i$. \\ Each differential component of $\boldsymbol{J_x}$ can be simply computed by differentiating the inverse transformation in~\eqref{hinv} with respect to each new latent component $\tilde{x}_i$

\begin{equation}
    \frac{\partial x_i}{\partial \tilde{x}_i} = \frac{\tilde{f}_i \Bigl ( \frac{\tilde{x}_i-\tilde{\mu}_i(\boldsymbol{\theta})}{\tilde{\sigma}_i(\boldsymbol{\theta})}\Bigr )}{\Phi^{-1} \Bigl [ \tilde{F}_i \Bigl ( \frac{\tilde{x}_i-\tilde{\mu}_i(\boldsymbol{\theta})}{\tilde{\sigma}_i(\boldsymbol{\theta})}\Bigr ) \Bigr ]}
    \label{difratio}
\end{equation}

\noindent
so that a complete representation of the Skew Gaussian Copula density can be obtained by putting together \eqref{fullden1} and \eqref{difratio} into the same equation

\begin{equation}
    \tilde{\pi}_{\SGC}(\boldsymbol{\tilde{x}} \vert \boldsymbol{\theta}, \boldsymbol{y}) = (2\pi)^{-\frac{N}{2}} \vert \boldsymbol{Q^*(\theta)} \vert^{\frac{1}{2}} \exp \Bigl [ -\frac{1}{2} [\boldsymbol{t(\tilde{x})}]^T \boldsymbol{Q^*(\theta)} [\boldsymbol{t(\tilde{x})}] \Bigr ] \prod_{i=1}^N \delta_i(\tilde{x}_i)
    \label{allden}
\end{equation}
with $\boldsymbol{t(\tilde{x})}=\boldsymbol{\Phi}^{-1} \Bigl [ \boldsymbol{\tilde{F}} \Bigl ( \frac{\boldsymbol{\tilde{x}}-\boldsymbol{\tilde{\mu}(\theta)}}{\boldsymbol{\tilde{\sigma}(\theta)}} \Bigr ) \Bigr ]\boldsymbol{\sigma(\theta)}$ and $\delta_i(\tilde{x}_i) = \frac{\tilde{f}_i \Bigl ( \frac{\tilde{x}_i-\tilde{\mu}_i(\boldsymbol{\theta})}{\tilde{\sigma}_i(\boldsymbol{\theta})}\Bigr )}{\Phi^{-1} \Bigl [ \tilde{F}_i \Bigl ( \frac{\tilde{x}_i-\tilde{\mu}_i(\boldsymbol{\theta})}{\tilde{\sigma}_i(\boldsymbol{\theta})}\Bigr ) \Bigr ] }$, $\forall i$.

\noindent
Easy to notice that the joint density in~\eqref{allden} is not properly Gaussian but still inherits the same precision structure from the Gaussian approximation in~\eqref{gau_dens} thanks to the Gaussian copula artifact. However, different choices of the transformation $\boldsymbol{h(\cdot)}$ in~\eqref{meancorr1} can still extract more useful approximations belonging to the same Skew Gaussian Copula class. \\ As an example, if the skewness is negligible than the transformation can just be assumed to be an identity so that $\boldsymbol{\tilde{z}}=\boldsymbol{z}$ and \eqref{meancorr1} is simplified into

\begin{equation}
    \boldsymbol{\tilde{x}}=\boldsymbol{\tilde{\sigma}(\theta)z}+\boldsymbol{\tilde{\mu}(\theta)}=\boldsymbol{x}-\boldsymbol{\mu(\theta)}+\boldsymbol{\tilde{\mu}(\theta)}
    \label{mcor}
\end{equation}
which can be referred as simple "mean correction" since it only applies location adjustments by shifting the original mean of the Gaussian approximation by the improved mean term $\boldsymbol{\tilde{\mu}(\theta)}$ of the Simplified Laplace marginals computed in INLA. Worth to point out that $\boldsymbol{\sigma(\theta)}$ and $\boldsymbol{\tilde{\sigma}(\theta)}$ are the same by construction of both the Gaussian and Simplified Laplace approximations. \\ Respectively the density in~\eqref{fullden1} would degenerate into a Gaussian Approximation with improved mean $\boldsymbol{\tilde{\mu}(\theta)}$ and precision matrix $\boldsymbol{Q^*(\theta)}$

\begin{equation}
    \tilde{\pi}_{\IG}(\boldsymbol{x} \vert \boldsymbol{\theta}, \boldsymbol{y})=(2\pi)^{-\frac{N}{2}} \vert \boldsymbol{Q^*(\theta)} \vert^{\frac{1}{2}} \exp \Bigl [ -\frac{1}{2} ( \boldsymbol{x}-\boldsymbol{\tilde{\mu}(\theta)})^T \boldsymbol{Q^*(\theta)} (\boldsymbol{x}-\boldsymbol{\tilde{\mu}(\theta)}) \Bigr ]
    \label{gau_mean}
\end{equation}
which we denote as \emph{Improved Gaussian Approximation}(IG). \\ Important to underline that both the Gaussian Approximation and its Improved version in~\eqref{gau_mean} belong to the Skew Gaussian Copula class: the improved case is clear while the standard case is obtained by assuming the set of cumulative marginals $\boldsymbol{F}$ for $\boldsymbol{\tilde{x}}$ to be exactly from the Gaussian approximation marginals $\tilde{\pi}_{\G}(x_i \vert \boldsymbol{\theta}, \boldsymbol{y})$, $\forall i$.

\subsection{Evaluating the corrections}

Since a proper representation of the transformed full conditional joint density derived from a combination of a Gaussian copula and a transformation is available in~\eqref{allden}, we can also precisely measure the location and skewness correction amount that arises after this transformation is applied. \\ Indeed, we remind from~\eqref{gau_dens} that the Gaussian approximation at the denominator evaluated at its mean $\boldsymbol{x} = \boldsymbol{\mu(\theta)}$ is

\begin{equation}
    \log \tilde{\pi}_{\G}(\boldsymbol{x} \vert \boldsymbol{\theta}, \boldsymbol{y}) \Bigl \vert_{\boldsymbol{x} = \boldsymbol{\mu(\theta)}}= \frac{1}{2} \vert \boldsymbol{Q^*(\theta)} \vert -\frac{N}{2} \log (2\pi)
    \label{gaeval}
\end{equation}

\noindent
Since our aim is to measure how much the correction amounts when we apply the new transformation on $\tilde{\pi}_G(\boldsymbol{x} \vert \boldsymbol{\theta}, \boldsymbol{y})$, we must evaluate the new class of joint approximations in~\eqref{allden} at the same mean point for the new latent random field $\boldsymbol{\tilde{x}}$. \\ Similarly to the notation applied in~\eqref{allden} we get the following

\begin{equation}
    \log \tilde{\pi}_{\SGC}(\boldsymbol{\tilde{x}} \vert \boldsymbol{\theta}, \boldsymbol{y}) \Bigl \vert_{\boldsymbol{\tilde{x}}= \boldsymbol{\mu(\theta)}} = \frac{1}{2} \log \vert \boldsymbol{Q^*(\theta)} \vert  -\frac{1}{2} [\boldsymbol{t(\mu(\theta))}]^T \boldsymbol{Q^*(\theta)} [[\boldsymbol{t(\mu(\theta))}]]+\sum_{i=1}^N \log \delta_i(\mu_i(\boldsymbol{\theta}))
    \label{difratioeval}
\end{equation}
Through the evaluated amounts in~\eqref{gaeval} and~\eqref{difratioeval}, we are able to account for the differential correction when using the new transformed joint density

\begin{align}
    \Delta_{\boldsymbol{x}, \boldsymbol{\tilde{x}}} & = \Bigl [ \log \tilde{\pi}_{\G}(\boldsymbol{x} \vert \boldsymbol{\theta}, \boldsymbol{y})-\log \tilde{\pi}_{\SGC}(\boldsymbol{\tilde{x}} \vert \boldsymbol{\theta}, \boldsymbol{y}) \Bigr ] \Bigl \vert_{(\boldsymbol{x}, \boldsymbol{\tilde{x}})= \boldsymbol{\mu(\theta)}} \nonumber \\
    &=  \frac{1}{2} [\boldsymbol{t(\mu(\theta))}]^T \boldsymbol{Q^*(\theta)} [[\boldsymbol{t(\mu(\theta))}]]-\sum_{i=1}^N \log \delta_i(\mu_i(\boldsymbol{\theta})) 
    \label{corrskeweval}
\end{align}
Trivial to notice that \eqref{corrskeweval} would degenerate into

\begin{equation}
    \Delta_{\boldsymbol{x}, \boldsymbol{\tilde{x}}} = \frac{1}{2}(\boldsymbol{\mu(\theta)}-\boldsymbol{\tilde{\mu}(\theta)})^T \boldsymbol{Q(\theta)}(\boldsymbol{\mu(\theta)}-\boldsymbol{\tilde{\mu}(\theta)})
\end{equation}
if we consider the Skew Gaussian Copula with mean corrected marginals in~\eqref{gau_mean} which, we remind, is a particular case of the class when the transformations $\boldsymbol{h(\cdot)}$ are assumed to be an identity.

\section{Joint Approximation for Linear Combination}
\label{sec:2}

The class of Skew Gaussian Copula joint approximations in Section \ref{sec:1}, can be used to construct an analytical approximation for joint densities  of the latent field components after integrating out the hyperparameter information $\boldsymbol{\theta}$. \\ In particular, we can get fast and quite accurate deterministic approximations for marginal densities of linear combinations within the latent field $\boldsymbol{x}$. The resulting multivariate results still belongs to the class of Skew Gaussian Copula introduced in Section \ref{sec:1} since we exploit moments up to order three of its respective full conditional densities depending on $\boldsymbol{\theta}$. A flawless derivation of these quantities allow to achieve a similar Gaussian Copula structure of the same class while using Skew Normal densities to approximate the marginals of the resulting multivariate object, which exactly correspond to each single linear combination by construction. \\ This alternative manipulation of the Skew Gaussian Copula enlarges the researcher toolbox for INLA by allowing more streamlined approximations and inference for additive linear combinations of the latent field. 

\subsection{Moments Matching}

Let us define a set of indexes $S= \{ i \vert i \in \{ 1,\dots, \vert N \vert\}, \vert \boldsymbol{x} \vert = N \}$ aiming to compute $\pi(\boldsymbol{x}_S \vert \boldsymbol{y})$ whose density can be decomposed into

\begin{equation}
    \pi(\boldsymbol{x}_S \vert \boldsymbol{y})= \int \pi(\boldsymbol{x}_S \vert \boldsymbol{\theta}, \boldsymbol{y}) \pi(\boldsymbol{\theta} \vert \boldsymbol{y}) \, d\boldsymbol{\theta}
    \label{submarg}
\end{equation}
Similarly to how we get Laplace approximations for the marginals in~\eqref{marg_lat}, we can construct an approximation for~\eqref{submarg} by exploiting the Skew Gaussian Copula of Section \ref{sec:1}

\begin{equation}
    \tilde{\pi}(\boldsymbol{x}_S \vert \boldsymbol{y}) \approx \sum_{k=1}^K \tilde{\pi}_{\SGC}(\boldsymbol{x}_S \vert \boldsymbol{\theta}_k, \boldsymbol{y}) w_k
    \label{subapprox}
\end{equation}
where $w_k=\tilde{\pi}(\boldsymbol{\theta}_k \vert \boldsymbol{y})\Delta_k$ such that $\sum_{k=1}^K w_k = 1$, that is, the weights are normalised. The joint densities \eqref{subapprox} are exactly the full conditional Skew Gaussian Copula approximations with corresponding density in~\eqref{allden}.  This class extends the Gaussian Approximation features by encoding skewness into its structure through the use of a Skew Normal transformation applied on the latent field. \\ A direct analytical approximation of~\eqref{subapprox} can be obtained by using a particular simpler option of the class, that is, the Improved Gaussian Approximation in~\eqref{gau_mean} which applies a correction to the mean while retaining its Gaussian structure. In other settings, this Gaussian representation may turn out to be a limit and that is why we want to exploit all the features of the Skew Gaussian Copula to model non skewed behaviour as well. However we must approach the approximation problem from a different way since this class does not have a proper shape. \\ Instead, we can take advantage of the mixture structure in~\eqref{subapprox} by computing moments up to order three to similarly reconstruct a Skew Gaussian Copula joint approximation to the entire sum. Basically, we integrate out the posterior summary information of  $\tilde{\pi}_{\SGC}(\boldsymbol{x}_S \vert \boldsymbol{\theta}_k, \boldsymbol{y})$ with respect to each integration point $\boldsymbol{\theta}_k$ and then combine them all into one general summary. Such manipulation leads to a new multivariate joint structure which still belongs to a Skew Gaussian Copula with new found moment information (this is going to be particularly useful later for linear combinations). \\ Therefore we can retain both the mean vector $\boldsymbol{\tilde{\mu}(\theta)}$ of the Skew Normal marginals and the covariance matrix structure $\boldsymbol{\Sigma^*_{SS}(\theta)}$ within a sub-block indexed by $S$. \\ Here the covariance matrix $\boldsymbol{\Sigma^*_{SS}(\theta)}$ represents the extracted solution within the assumed subset $S$ of the linear system $\boldsymbol{Q\Sigma}= \boldsymbol{I}$. \\ Since $\boldsymbol{x}_S \vert \boldsymbol{y} \sim \tilde{\pi}(\boldsymbol{x}_S \vert \boldsymbol{y})$ by assumption, then we can calculate its moments in terms of a Skew Gaussian Copula as follows

\begin{align}
    \E_{\SGC}[\boldsymbol{x}^{p}_S \vert \boldsymbol{y}] &= \int_{S} \boldsymbol{x}^{p}_S \tilde{\pi}(\boldsymbol{x}_S \vert \boldsymbol{y}) \, d \boldsymbol{x}_S \nonumber \\
    &= \int_{S} \boldsymbol{x}^{p}_S \sum_{k=1}^K \tilde{\pi}_{\SGC}(\boldsymbol{x}_S \vert \boldsymbol{\theta}_k, \boldsymbol{y}) w_k \, d \boldsymbol{x}_S \nonumber \\
    &= \sum_{k=1}^K w_k \int_{S} \boldsymbol{x}^{p}_S \tilde{\pi}_{\SGC}(\boldsymbol{x}_S \vert \boldsymbol{\theta}_k, \boldsymbol{y}) \, d \boldsymbol{x}_S \nonumber \\
    &= \sum_{k=1}^K w_k \E_{\SGC}[\boldsymbol{x}^{p}_S \vert \boldsymbol{\theta}_k, \boldsymbol{y}] \nonumber \\
    &= \sum_{k=1}^K w_k \boldsymbol{\tilde{\mu}}^{(p)}_S(\boldsymbol{\theta}_k)
    \label{det_mom}
\end{align}
where $\boldsymbol{\tilde{\mu}}^{(j)}_S(\boldsymbol{\theta}_k)$ defines all the $j^{th}$ marginal non central moments of the full conditional densities in~\eqref{det_mom}. The moments up to order $p=3$ allows to construct a Skew Gaussian Copula joint approximation to $\pi(\boldsymbol{x}_S \vert \boldsymbol{y})$ with mean $\E_{\SGC}[\boldsymbol{x}_S \vert \boldsymbol{y}]=\boldsymbol{\tilde{\mu}}_S$, covariance structure $\boldsymbol{\Sigma}^*$ and Skew Normal marginals by matching the third order moment $\E_{\SGC}[\boldsymbol{x}^3_S \vert \boldsymbol{y}]$. Through the third order moment we can get the the skewness $\gamma_{\SGC}(\boldsymbol{x}_S \vert \boldsymbol{y})$ by using the general formula 

\begin{equation}
    \gamma_{\SGC}(\boldsymbol{x} \vert \boldsymbol{y})  =  \frac{\E_{\SGC}(\boldsymbol{x}_S^3 \vert \boldsymbol{y})-3\E_{\SGC}(\boldsymbol{x}_S^2 \vert \boldsymbol{y})\E_{\SGC}(\boldsymbol{x}_S \vert \boldsymbol{y})+2\E_{\SGC}^3(\boldsymbol{x}_S \vert \boldsymbol{y})}{[\E_{\SGC}(\boldsymbol{x}_S^2 \vert \boldsymbol{y})-\E_{\SGC}^2(\boldsymbol{x}_S \vert \boldsymbol{y})]^{\frac{3}{2}}}
    \label{skewform}
\end{equation}
As soon as mean, variance and skewness for each marginal is available, we can derive proper Skew Normal densities by simply matching their respective moments. Indeed, we can map each skewness component to the respective Skew Normal parameters given by location $\xi$, scale $\omega$ and skewness index $\alpha$. This is accomplished by using the method of moment estimation procedure (MME), which sets a three equations linear system after matching the first three order moments (\cite{ghorbanzadeh_2017}). \\ The final results are obtained by an available straightforward parameterization derived from the Skew Normal family

\begin{definition}[\textbf{$\delta$-parameterization}]
Let $\{ \xi_i, \omega_i, \alpha_i \}$ be a set of parameters triplet of a Skew Normal distribution and $\gamma_i$ the skewness for each marginal latent field term $x_i$ with mean $\mu_i$ and variance $\sigma_i^2$ for $i=1,\dots,N$. Then we can analytically compute a new $\tilde{\delta}_i$ parameter 
\begin{equation}
\tilde{\delta}_i = \sign(\gamma_i) \sqrt{\frac{ \frac{\pi}{2} |\gamma_i|^{2/3}}{\Bigl (\frac{4-\pi}{2} \Bigr)^{2/3}+|\gamma_i|^{2/3}}}
\end{equation}
in terms of the skewness $\gamma_i$. Based on MME, we get
\begin{align*}
\tilde{\alpha}_i &= \frac{\tilde{\delta}_i}{\sqrt{1-\tilde{\delta}_i^2}} \\
\tilde{\omega}_i &= \sqrt{\frac{\pi \sigma_i^2}{\pi-2\tilde{\delta}_i^2}} \\
\tilde{\xi}_i &= \mu_i-\tilde{\omega}_i \tilde{\delta}_i \sqrt{\frac{2}{\pi}}
\end{align*}
which is the $\delta$-parameterization for the initial triplet.
\label{deltapar}
\end{definition}

\noindent
Each marginal of the Skew Gaussian Copula joint approximation $\tilde{\pi}_{\SGC}(\boldsymbol{x}_S \vert \boldsymbol{y})$ is then represented by a Skew Normal density derived from its respective linear system resolution. \\ \\ In this framework, we are also able to approximate additive linear combinations of the latent field since their density assumptions tend to be more Gaussian like as the dimension $\vert S \vert$ increases, meaning that the Skew Gaussian Copula class already represents a good enough approximation candidate. \\  Similarly to~\eqref{subapprox}, we can define an additive linear combination $l(\boldsymbol{x})= \boldsymbol{Ax}$ as a vector of dimension $M$ with $\boldsymbol{A}$ being a $M \times N$ matrix of indexes that generalizes the notation $S$ into $M$ additive linear combinations and $\boldsymbol{x} $ follows the distribution of a Skew Gaussian Copula as in~\eqref{allden}. \\ Its joint density can then be approximated by

\begin{equation}
    \tilde{\pi}(\boldsymbol{Ax} \vert \boldsymbol{y}) \approx \sum_{k=1}^K \tilde{\pi}_{\SGC}(\boldsymbol{Ax} \vert \boldsymbol{\theta}_k, \boldsymbol{y}) w_k
    \label{subapproxlin}
\end{equation}
with Skew Gaussian Copula approximations applied on the linear combination vector $\boldsymbol{Ax}$. \\ Following similar steps carried out in~\eqref{det_mom} we can compute the moments for~\eqref{subapproxlin} by using

\begin{equation}
    \E_{\SGC}[(\boldsymbol{Ax})^{p} \vert \boldsymbol{y}] = \sum_{k=1}^K
w_k \E_{\SGC}[(\boldsymbol{Ax})^{p} \vert \boldsymbol{\theta}_k, \boldsymbol{y}]
\label{det_lin_mom}
\end{equation}
where $(\boldsymbol{Ax})^{p}=\boldsymbol{A}^p\boldsymbol{x}^p$ in \eqref{det_lin_mom} denotes a power vector-wise evaluation of each component in the $M$-dimensional linear combination vector $\boldsymbol{Ax}$. Therefore we can compute its respective posterior moments up to order $p=3$ as follows 

\begin{align}
    \E_{\SGC}[\boldsymbol{Ax} \vert  \boldsymbol{y}] &= \boldsymbol{A} \sum_{k=1}^K w_k \E_{\SGC}[\boldsymbol{x} \vert \boldsymbol{\theta}_k, \boldsymbol{y}]= \boldsymbol{A} \sum_{k=1}^K w_k \boldsymbol{\tilde{\mu}(\theta_k)}= \boldsymbol{A} \boldsymbol{\tilde{\mu}} \nonumber \\
    \E_{\SGC}[(\boldsymbol{Ax})^2 \vert  \boldsymbol{y}] &= \sum_{k=1}^K w_k \V_{\SGC} (\boldsymbol{Ax} \vert \boldsymbol{\theta}_k, \boldsymbol{y}) +\sum_{k=1}^K w_k [\E_{\SGC}[\boldsymbol{Ax} \vert \boldsymbol{\theta}_k, \boldsymbol{y}]]^2 \nonumber \\ 
    &= \diag [\boldsymbol{A}\boldsymbol{\Sigma}^*\boldsymbol{A}^T]+[\boldsymbol{A}\boldsymbol{\tilde{\mu}}]^2 
    \label{momtwo}
\end{align}
where $\boldsymbol{\tilde{\mu}}$ and $\boldsymbol{\Sigma}^*$ are respectively the mean vector and covariance matrix of the Skew Gaussian Copula joint approximation after integrating out $\boldsymbol{\theta}$. Since we only need to evaluate the third moment to get the skewness of the linear combination $\boldsymbol{Ax}$, then we can simply use its central moment representation

\begin{align}
    \E_{\SGC}[(\boldsymbol{Ax}-\boldsymbol{A\tilde{\mu}})^3 \vert  \boldsymbol{y}] &= \sum_{k=1}^K w_k \E_{\SGC}[(\boldsymbol{Ax}-\boldsymbol{A\tilde{\mu}(\theta_k)})^3 \vert \boldsymbol{\theta}_k, \boldsymbol{y}] \nonumber \\
    &= \boldsymbol{A}^3 \E_{\SGC}[(\boldsymbol{x}-\boldsymbol{\tilde{\mu}})^3 \vert \boldsymbol{y}] \nonumber \\
    &= \boldsymbol{A}^3 \boldsymbol{\gamma}_{\SGC}(\boldsymbol{x} \vert \boldsymbol{y})[\diag (\boldsymbol{\Sigma^*})]^{\frac{3}{2}}
    \label{threecentral}
\end{align}
which only depends on the central third order moments of the latent field $\boldsymbol{x}$ since all the other mixed moments are zero (see \cite{sym2010} for more details on moments of a multivariate Gaussian distribution). Thus, the overall skewness for the linear combination vector $\boldsymbol{Ax}$ can easily be evaluated as

\begin{equation}
    \boldsymbol{\gamma}_{\SGC}(\boldsymbol{Ax} \vert \boldsymbol{y})= \frac{\E_{\SGC}[(\boldsymbol{Ax}-\boldsymbol{A\tilde{\mu}})^3 \vert  \boldsymbol{y}]}{(\diag [\boldsymbol{A}\boldsymbol{\Sigma}^*\boldsymbol{A}^T])^\frac{3}{2}}
\end{equation}
Since the Gaussian copula structure remains unscathed except for the marginals, the same argument of~\eqref{subapprox} applies to the Skew Gaussian Copula in~\eqref{subapproxlin} so that the first two order non central moments in~\eqref{momtwo} provide information for the mean summary $\boldsymbol{A}\boldsymbol{\tilde{\mu}}$ and the new covariance structure $\boldsymbol{A}\boldsymbol{\Sigma}^*\boldsymbol{A}^T$ where $\boldsymbol{\Sigma}^*$ is the entire $N \times N$ covariance structure of the latent field $\boldsymbol{x}$. The third central order moment in~\eqref{threecentral} is not zero by Skew Gaussian Copula construction and allows to encode and model skewness into each linear combination marginal specified by $\boldsymbol{Ax}$. \\ Cross covariance terms within $\boldsymbol{A}\boldsymbol{\Sigma}^*\boldsymbol{A}^T$ relates to the covariance amongst different linear combinations $h$ and $l$ and can be computed as follows

\begin{align}
    \Cov_{\SGC} \Bigl [ \sum_{i=1}^N A_{h,i} x_i, \sum_{i=1}^N A_{l,i} x_i \Bigl \vert \boldsymbol{y}) \Bigr ] &= \E_{\SGC} \Bigl [\sum_{i=1}^N A_{h,i} x_i \sum_{i=1}^N A_{l,i} x_i \Bigl \vert \boldsymbol{y}) \Bigr ]- \nonumber \\
    &- \E_{\SGC} \Bigl [\sum_{i=1}^N A_{h,i} x_i \Bigl \vert \boldsymbol{y}) \Bigr ] \E_{\SGC} \Bigl[ \sum_{i=1}^N A_{l,j} x_i \Bigl \vert  \boldsymbol{y}) \Bigr ] \nonumber \\
    &= \sum_{i=1}^N A_{h,i}A_{l,i} \Sigma^*_{i,i}+ \sum_{i=1}^N \sum_{j=1}^{i-1} (A_{h,i}A_{l,j}+A_{h,j}A_{l,i}) \Sigma^*_{i,j} 
    \label{covdetlin}
\end{align}
As soon as all the required moments exist and are available, we can construct a simpler but accurate approximation to the joint density $\pi(\boldsymbol{Ax} \vert \boldsymbol{y})$ which still belongs to the Skew Gaussian Copula class. Indeed we can write $\pi(\boldsymbol{Ax} \vert \boldsymbol{y}) \approx \tilde{\pi}_{\SGC}(\boldsymbol{Ax} \vert \boldsymbol{y})$ and the same applies for $\pi(\boldsymbol{x}_S \vert \boldsymbol{y})$. \\ The resulting joint approximations inherit the same properties of the Skew Gaussian Copula full conditional approximations while integrating out $\boldsymbol{\theta}$. \\ The posterior summaries of the above approximations are then derived from the weighted moment structure in~\eqref{det_mom} and~\eqref{det_lin_mom}. \\ \\ In order to understand how we can construct these Skew Gaussian Copula approximations $\tilde{\pi}_{\SGC}(\boldsymbol{x} \vert \boldsymbol{y})$ and $\tilde{\pi}_{\SGC}(\boldsymbol{Ax} \vert \boldsymbol{y})$,  let us consider an example where we have a reduced latent field $\boldsymbol{x}=(x_1, x_2)$, corresponding to $S=2$, and matrix of indexes $\boldsymbol{A}=\begin{pmatrix}
1 & 1 \\ 
1 & -1
\end{pmatrix}$ to define the linear combinations $\boldsymbol{Ax}=(x_1+x_2, x_1-x_2)$. Based on the formulas in~\eqref{det_mom}, we assume that $\pi(\boldsymbol{x} \vert \boldsymbol{y})$ is approximated by a Skew Gaussian Copula with $\E_{\SGC}(\boldsymbol{x} \vert \boldsymbol{y})=(\boldsymbol{\tilde{\mu}}_1, \boldsymbol{\tilde{\mu}}_1)=(1,2)$, $\boldsymbol{\Sigma}^* = \begin{pmatrix}
2 & 1 \\ 
1 & 5
\end{pmatrix}$ and skewness $\boldsymbol{\gamma}_{\SGC}(\boldsymbol{x} \vert \boldsymbol{y})=(-0.4,0.6)$. Then we aim to construct a Skew Gaussian Copula approximation to $\pi(\boldsymbol{Ax} \vert \boldsymbol{y})$ with generic observations $\boldsymbol{y}$. \\ From~\eqref{momtwo} and~\eqref{threecentral} we compute all the information we need 

\begin{align}
    \E_{\SGC}[\boldsymbol{Ax} \vert \boldsymbol{y}] &= \boldsymbol{A\tilde{\mu}}=(\tilde{\mu}_1+\tilde{\mu}_2, \tilde{\mu}_1-\tilde{\mu}_2)=(3,-1) \nonumber \\
    \boldsymbol{A}\boldsymbol{\Sigma}^*\boldsymbol{A}^T &= \begin{pmatrix}
9 & -3 \\ 
-3 & 5
\end{pmatrix} \nonumber \\
    \boldsymbol{\gamma}_{\SGC}[\boldsymbol{Ax} \vert \boldsymbol{y}] &= \frac{\boldsymbol{A}^3 \boldsymbol{\gamma}_{\SGC}(\boldsymbol{x} \vert \boldsymbol{y})[\diag (\boldsymbol{\Sigma^*})]^{\frac{3}{2}}}{(\diag [\boldsymbol{A}\boldsymbol{\Sigma}^*\boldsymbol{A}^T])^\frac{3}{2}}= (0.206, -0.701)
    \label{ex_momdet}
\end{align}
By construction of the Skew Gaussian Copula object, the corresponding marginals extrapolated from $\boldsymbol{Ax}$ denoted by $\pi(x_1+x_2 \vert \boldsymbol{y})$ and $\pi(x_1-x_2 \vert \boldsymbol{y})$ are approximated by Skew Normal densities with respective location $\xi$, scale $\omega$ and skewness parameter $\alpha$ 

\begin{align}
    \tilde{\pi}(x_1+x_2 \vert \boldsymbol{y}) &\approx SN(-0.107, 1.796, 1.217) \nonumber \\
    \tilde{\pi}(x_1-x_2 \vert \boldsymbol{y}) &\approx SN(4.633, 3.454, -3.233)
    \label{sn_marg_ex}
\end{align}
obtained by matching the moments in~\eqref{ex_momdet} with respect to the $\delta$-parameterization in Definition \ref{deltapar}. These skewed marginals provide straightforward analytical approximations for each additive linear combination component in $\boldsymbol{Ax}$, therefore easing their inference analysis. \\ Although this additional joint approximation may lack accuracy in more complex settings, the available results are convincing and allow streamlined and fast marginal inference for additive linear combinations of the latent field (see Appendix A for a practical example).

\section{Mixture of Skew Gaussian Copula densities} 
\label{sec:3}

Up to this point two factors are certain about INLA methodology and applicability: marginal posterior inference is well established by using Laplace approximations while joint posterior inference can only easily be carried out for additive linear combinations. For more complex and not well behaved cases we may need to switch to more proper and accurate joint approximations which can actually be obtained with a sampling Monte Carlo scheme. \\ \\ Unlike sampling based methods such as Markov Chain Monte Carlo, INLA is unable to provide accurate posterior results from a joint density since its shape is unknown and no deterministic approximation is possible in general terms. A rare exception is represented by the new deterministic artifact shown in Section \ref{sec:2} which allows quite accurate analytical approximation for linear combinations and functionals within a certain subset of the joint parameter space. \\ However, in the context of Latent Gaussian Models, we are able to construct a quite accurate approximation. \\ \\ By recalling a similar structure in~\eqref{marg_lat}  and approximation strategies used in~\eqref{marg_inla}, we can use the new defined class of Skew Gaussian Copula densities $\tilde{\pi}_{\SGC}(\boldsymbol{\tilde{x}} \vert \boldsymbol{\theta}, \boldsymbol{y})$ from Section \ref{sec:1} to get an approximation for the overall joint $\pi(\boldsymbol{x}, \boldsymbol{\theta} \vert \boldsymbol{y})$. \\ Similarly to how INLA approximates marginal in~\eqref{marg_inla}, the respective joint posterior approximation can be written down as follows

\begin{equation}
\tilde{\pi}(\boldsymbol{\tilde{x}},\boldsymbol{\theta}|\boldsymbol{y}) \propto \sum_k \tilde{\pi}_{\SGC} (\boldsymbol{\tilde{x}}|\boldsymbol{\theta},\boldsymbol{y}) \tilde{\pi}(\boldsymbol{\theta}|\boldsymbol{y}) \mathbb{1}_{[\boldsymbol{\theta}=\theta_k]} \Delta_k
\label{joint:approx}
\end{equation}
with $\{ \boldsymbol{\theta}_k, k=1,\dots,K \}$ being the configuration points obtained in the grid exploration phase and $\Delta_k$ the usual area weights. The approximation \eqref{joint:approx} can be interpreted as a \emph{mixture of Skew Gaussian Copula distributions} since these densities play the main role. \\ Indeed, this class of joint approximations represents the only source of contribution and error to the full joint approximation accuracy. \\ \\ As similar as it may appear, the approximation in~\eqref{joint:approx} does not carry the same accuracy as the posterior marginals $\tilde{\pi}(x_i \vert \boldsymbol{y})$ since it lacks a Laplace approximation step. The mixture structure of \eqref{joint:approx} easily allow to achieve an approximation to the true density in~\eqref{joint:eq} by using a sampling Monte Carlo approach on the pre-computed grid points of the hyperparameter space, as a result of the internal joint approximations obtained in~\eqref{joint:hyper}. \\ The sampling scheme can be summarised in two simple steps:

\begin{description}
\item[$\bullet$] we draw samples from the entire hyperparameter set $\boldsymbol{\theta}$ using the pre-computed configuration points $\{ \boldsymbol{\theta}_k, k=1,\dots,K\}$ for exploring~\eqref{joint:hyper} in terms of their mass probability function, as shown in~\eqref{joint:approx};
\item[$\bullet$] for each sampled configuration point $\boldsymbol{\theta}_k$ a $N$-dimensional sample is drawn from $\tilde{\pi}_{\SGC} (\boldsymbol{x} \vert \boldsymbol{\theta},\boldsymbol{y})$ with weighted probability $\tilde{\pi}(\boldsymbol{\theta}|\boldsymbol{y}) \mathbb{1}_{[\boldsymbol{\theta}=\theta_k]}$ that sum up to one $\forall k$.
\end{description}

\noindent
The new Skew Gaussian Copula class turns out to be the key role for both computing and improving the joint posterior approximation in~\eqref{joint:approx}. We can always get samples from this multivariate density no matter the model to fit, and we can do it fast.

\subsection{Skew Normal Marginal Transformations}
\label{sec:1.1}

The joint Gaussian approximation used in~\eqref{joint:hyper}, which is used for the Gaussian approximation strategy as well, represents the most accurate and fast inferential tool adopted by INLA when the likelihood model is Gaussian and does not include large anti-symmetric deviations. \\ As described in Section \ref{sec:1}, we formally established that this joint approximation may be lackey for properly "capturing" extreme skewed outcomes derived from GLMM models with Poisson or Binomial likelihood assumptions (see also \cite{ferkingstad2015latent}). Hence we construct and define a more general class which combines a Gaussian copula structure with skewed transformed marginals: the Skew Gaussian Copula class. \\ If the transformation is an identity then the resulting joint density is the Improved Gaussian Approximation in~\eqref{gau_mean} where the latent field marginals are still Gaussian but shifted in their means by well defined correction terms. This correction is referred as mean correction~\eqref{mcor} and amounts to the respective Skew Normal mean summaries coming from the Laplace approximation strategies. \\ This improved mean corrected version of the joint Gaussian approximation has been internally implemented in INLA being the default option for years by exploiting the GMRFs sampling features. \\ The Skew Gaussian Copula class of joint densities gets general as soon as we exploit the Skew Normal transformation in~\eqref{meancorr1} which applies marginal corrections on both location and skewness: we refer to this correction as skewness correction. \\ Although this generalized correction may bring more accuracy on a larger scale of applications, its implementation turns out to be way more cumbersome and not computational friendly. \\ \\ Most of the extreme modeling cases can be accurately described by using the mean correction only in~\eqref{mcor} but this one alone does not resolve all possibilities.  \\ Conceptually, the large deviation is spotted by observing high skewness outcomes in the posterior marginals of the resulting joint object. As a result,  the approximation in~\eqref{joint:approx} is always the most accurate when the generalized skewness correction in~\eqref{meancorr1} is employed.

\subsubsection{Mean and Skewness correction comparison}

\begin{figure}[hbt!]
  \centering
  \includegraphics[scale=0.8]{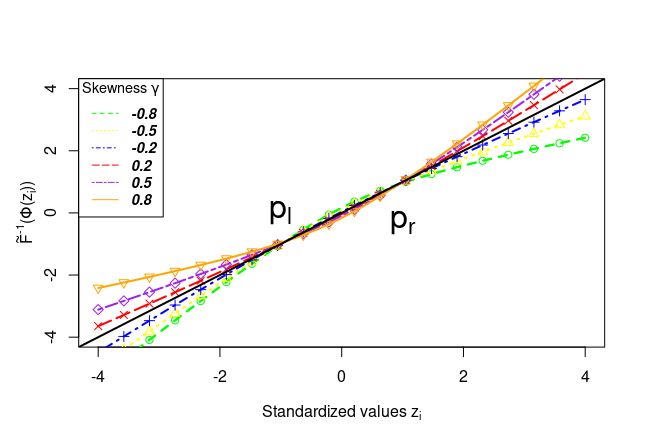}
\caption{Standardized latent values $z_i$ compared with the skewness corrected values $\tilde{x}_i$ through the quantile Skew Normal transformation $\tilde{F}^{-1}(\cdot)$ on the range $(-4,4)$. The intersection points $p_l$ and $p_r$ are used as an exact threshold for detecting the skewness effect in the tails of the $i^{th}$ marginal distribution.}
\label{skew:plot}       
\end{figure}

Differences between mean~\eqref{mcor} and skewness~\eqref{meancorr1} correction can be observed in Figure \ref{skew:plot} where several levels of skewness are assumed on the Skew Normal transformed latent marginals compared to the non transformed latent terms $\{ x_1, \dots, x_N\}$ where the transformation is an identity, meaning that input and output are the same. \\ \\

\noindent
The straight 45 degrees black line represents the mean corrected standardized values $z_i$ while the colored lines show the skewness correction under the quantile Skew-Normal transformation $\tilde{F}^{-1}()$ as it applies in~\eqref{meancorr1}. \\ The intersection points $p_l$ and $p_r$ just underline a marginal threshold to detect when the skewness effect triggers a change in the distribution. Easy to see that almost no changes happen in the range $[p_l,p_r]$, which is approximately equal to $\approx [-1.5,1.5]$.  \\ The mean correction described by the black line is straight simply because no transformation is applied in this case; that is, the plotted values on the range are the same. Instead, the skewness correction colored lines underline different behaviors according to the different assumed skewness values shown in the legend on the left. \\ Here six different levels of skewness from -0.8 to 0.8 have been chosen, and the pattern is straightforward to interpret: higher is the skewness, higher is the effect on the marginal latent distribution with a greater focus on the tails. Moreover, a positive (negative) skewness effect leads to mostly underestimate (overestimate) the estimated latent variables when the mean corrected Gaussian approximation is used. \\ As an example, if we evaluate $z_i=3$ under marginal skewness equal to -0.8, we would end up with a corrected result below 2 instead of 3 by mean correction assumption. However, summaries like mean or median may not be able to detect this effect when it is there since most of the correction effect kicks in the tails of the distribution. \\ It should be clear by now that the deviation can be severe if the marginal skewness effect is reasonably significant, and the plot confirms this hypothesis. \\The computations of the transformation $\tilde{F}^{-1}()$ for the skewness correction are not straightforward and can require many evaluations proportional to the dimension of the latent field. As mentioned earlier,  a direct computation may be too heavy so that a faster approach may be required.

\subsubsection{Mapping Strategy for fast computations}

A direct approach to apply the skewness correction~\eqref{meancorr1} onto the latent field variables to construct a Skew Gaussian Copula joint density requires solving the following vectorial quantile equation many times

\begin{equation}
\tilde{F}^{-1}[\Phi(\boldsymbol{x})]=\boldsymbol{p}
\label{quantile:eq}
\end{equation}

\noindent
since the resulting approximation need to be evaluated for each configuration point $\boldsymbol{\theta}_k$ as stated in~\eqref{joint:approx} for getting a joint posterior approximation of a Latent Gaussian Model. \\ The number of operations are then proportional to $N \cdot s$ where $N = n+n_p$ is the dimension of the latent field $\boldsymbol{x}$ while $s$ is the pre-defined number of samples. This number can increase fast if both $N$ and $s$ are high. \\ \\ Therefore we must solve the equation efficiently in order to avoid unnecessary computational burden. Although the quantile equation in~\eqref{quantile:eq} does not have a closed-form solution, it is possible to compute exact evaluations through an optimization step. \\ In R language, an implementation of the solver is encoded in the function \texttt{qsn()} within the Skew Normal package \texttt{sn} and mainly use Newton Raphson steps or finite difference methods if the default solver fails. The combination of Newton Raphson steps and evaluations of the Gaussian cumulative densities $\Phi(\cdot)$ can be heavy if $N \cdot s$ dimension is high. Also, the lack of vectorization for different skewness values is as critical as the slow optimization solver. Not even the multivariate version of the functions, as suggested by \cite{Azzalini_1999} can fix this issue since the required operations are marginal and independent, and more, a multivariate quantile version does not exist. \\ Instead, the entire evaluation process can be handled through an automatic and fast process by tabulating all the solutions and then combine them in a smart way while avoiding unnecessary computations (see Appendix C).\\ The $\delta$-parameterization in Definition \ref{deltapar} is fast and accurate for the required computational standards. This new approach is summarized in Proposition \ref{prop}.

\begin{proposition}\label{prop}[\textbf{Mapping and Interpolation two-way strategy}]
\\
Let us assume to have access to all marginal skewness $\{ \gamma_i, i=1,\dots, N\}$ related to the latent field $\boldsymbol{x}= \{ x_i, i=1,\dots, N \}$ and define the Skew Normal mapping function according to Definition \ref{deltapar}. Then 

\begin{enumerate}
\item First a local cache of object files is being created within the local private INLA environment in R by assuming few useful initial constants;
\item Next the $\delta$-parameterization is exploited to map each skewness $\gamma_i$ into the respective Skew Normal triplet of parameters $\{\xi_i, \omega_i, \alpha_i \}$. These results are used to fit the correct marginal Skew Normal distributions for each $x_i$;
\item For each Skew-Normal marginal a fixed number of points are pre-computed and stored. Then they are used in an interpolation process to get all possible solutions;
\item Finally, for each marginal skewness,  the correct interpolant is detected by using binary search algorithm. 
\end{enumerate}
\label{map2d}
\end{proposition}

\noindent
By this procedure, the vectorial quantile equation in~\eqref{quantile:eq} can be solved from a two-way dimensional perspective surpassing the limits of the available methodologies where the input is a $N \times s$ dimensional object.

\section{Numerical simulations}
\label{sec:4}

This last section aims to explore the advantages of applying the new built in implementations of the Skew Gaussian Copula class object into R-INLA within a joint inference framework. Some examples and application of the mean corrected version of this class introduced in Section \ref{sec:1} appear in \cite{seppa2018} or \cite{wakefield_2016_spatial}. \\ The skewness correction effect of the Skew Gaussian Copula class is more general and grants some accuracy improvements for the joint inference analysis at the cost of additional computations that are negligible in most of the settings. In order to detect the advantages of utilizing the more general version of the class, we construct some extreme examples by simulating data from both a Poisson and Binomial likelihoods. This choice leads to posterior marginal representations of the joint result that are heavily skewed. Also, the resulting skewness scheme is dense so that all posterior marginals obtained from each grid point of the hyperparameter space are skewed, therefore adding computations to the new skewness corrected implementation. All the mean and skewness corrected marginal results from INLA are then compared to their MCMC counterpart with the exact same simulation settings carried out in JAGS ( \cite{plummer_2003_dsc}). As soon as the accurate implementation of the Skew Gaussian Copula object in all its different marginal corrections is verified, we provide examples for the manipulation of the same into the alternative joint posterior inference analysis for additive linear combinations described in Section \ref{sec:2}. A detailed overview in R of these new tools can be checked on Appendix A.

\subsection{Posterior marginal corrected inference}
\label{sec:4.1}

In order to show the features of the Skew Gaussian Copula class when constructing a joint posterior approximation of the model, we set a comparative Bayesian analysis using both MCMC methods, through JAGS software, and R-INLA Laplace strategies. The applications are based on data simulations from Poisson and Binomial likelihoods within a hierarchical GLMM model framework. \\ The hierarchical structure of the Poisson example is described as follows

\begin{gather*}
\boldsymbol{y}|\boldsymbol{\mu} \sim \text{Poi}(\boldsymbol{\mu}) \\
\boldsymbol{\mu} = \exp(\alpha + \boldsymbol{u}) \\
\boldsymbol{u} \sim \text{N}_m (\boldsymbol{0}, \tau^{-1} \boldsymbol{I}) \\
\alpha \sim \text{N}(0,1000) \\
\tau \sim \text{Ga}(0.1,0.1)
\end{gather*}

\noindent
The data $\boldsymbol{y}$ have been simulated from a Poisson distribution with $N=50$ observations and $M=10$ randomized groups for the vector of random effect $\boldsymbol{u}$ and each of these was then simulated from a Normal distribution with standard deviation $\sigma = \sqrt{\tau^{-1}}= 1.5$ to trigger high posterior marginal skewness results. Similarly, the Binomial example is constructed by using a logit link function applied to a generic probability parameter $p$. \\ These simulation settings lead to posterior marginals of the latent field to be quite skewed, and therefore we expect the mean correction version of the Skew Gaussian Copula to be inaccurate to properly retrieve the true results. In this Bayesian framework, it is well known that MCMC methods require long-run simulations to be reliable in terms of the Monte Carlo error. Instead, INLA relies on deterministic marginal posterior approximations which are empirically accurate. However, as stated in Section \ref{sec:3}, a full joint posterior inference of a LGM model is only attainable through a sampling based approach in terms of a mixture of Skew Gaussian Copula densities. The resulting joint posterior approximation does benefit from the marginal corrections derived from the Skew Gaussian Copula class. \\ The computational setup in the R language for both software is reported below

\begin{description}
\item[$\bullet$ \textbf{JAGS}:] simulating $6 \times 10^6$ samples for 20 independent Markov Chains with $10^2$ iterations thinning and dropping $10^6$ samples as burn-in. We handled the simulation in parallel by using our computational server at KAUST whose specs are: Intel(R) Xeon(R) Gold 6130 CPUs@2.10GHz with 512 Gb of RAM, two sockets with 16 cores each with two threads per core. This parallel setting takes around 11 minutes to complete.
\item[$\bullet$ \textbf{INLA}:] simulating $10^5$ samples from the joint sampler both with mean and skewness correction. The mean corrected strategy takes around 4 seconds while the skewness corrected one takes around 14 seconds on average. By using $10^4$ samples the computational times are respectively 0,6 and 1,5 seconds for both mean and skewness corrections. \\ All present work time results have been run on a Dell laptop whose specs are: Intel(R) Core(TM) i7-10710U CPU @ 1.10GHz
with 16 Gb of RAM, one socket with 6 cores with two threads per core. \\ Important to underline that this scheme is computationally heavy for the skewness correction since a large number of evaluations are needed. In most of the settings there is no need to use so many samples (because of the way the joint sampler is constructed)  and way less skewed marginals are observed. In these cases the computational difference amongst the two corrections is almost negligible. \\ This average time is obtained under 100 function replications. Appendix B shows that the slow down would be way heavier if we used, for example, default R approaches instead of our alternative strategy in \ref{map2d}.
\end{description}

\begin{figure}[bp!]
  \centering
  \includegraphics[scale = 0.7]{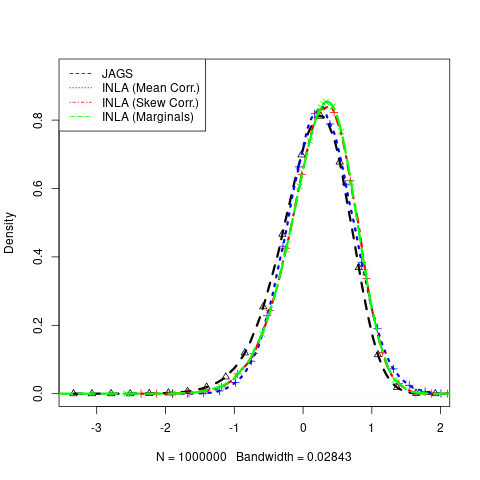}
\caption{Posterior marginal representation for linear predictor $\eta_9$ of the Poisson GLMM model with marginal skewness is around -0.38 for all the configuration points. The curves display the outcomes from different strategies: posterior marginal from JAGS (black), mean corrected (blue) and skewness corrected (red) marginal from the SGC and the Simplified Laplace posterior marginal (green) computed by INLA. }
\label{psla9}      
\end{figure}

\begin{figure}[bp!]
  \centering
  \includegraphics[scale = 0.7]{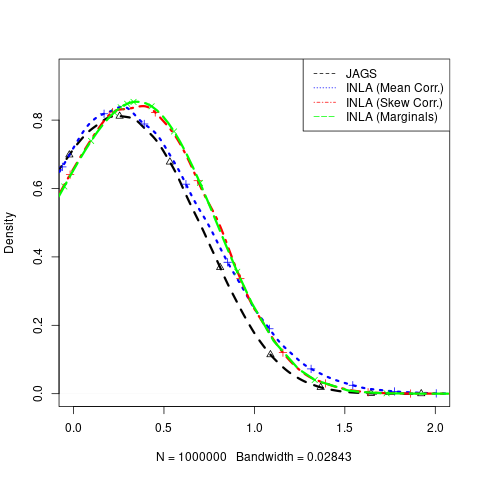}
\caption {A focus on the right tail of the linear predictor $\eta_9$ of the Poisson GLMM model where the skewness propagation to the tail is more evident. The skewness corrected marginal (red) from the SGC totally matches with the Simplified Laplace marginal result (green).}
\label{pslat9}       
\end{figure}

\begin{figure}[bp!]
  \centering
  \includegraphics[scale = 0.7]{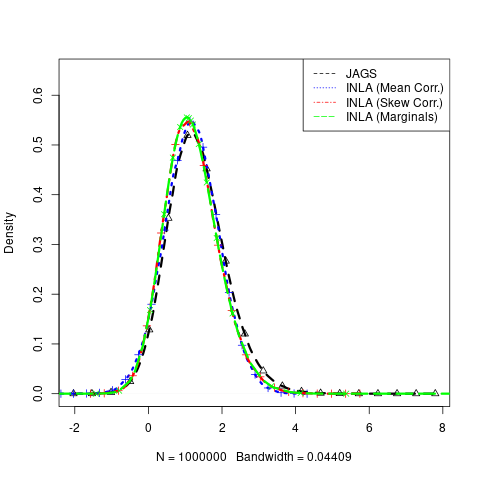}
\caption{Posterior marginal representation for linear predictor $\eta_{14}$ of the Binomial GLMM model with marginal skewness is around 0.38 for all the configuration points. The curves display the outcomes from different strategies: posterior marginal from JAGS (black), mean corrected (blue) and skewness corrected (red) marginal from the SGC and the Simplified Laplace posterior marginal (green) computed by INLA.  }
\label{bsla14}       
\end{figure}

\begin{figure}[bp!]
  \centering
  \includegraphics[scale = 0.7]{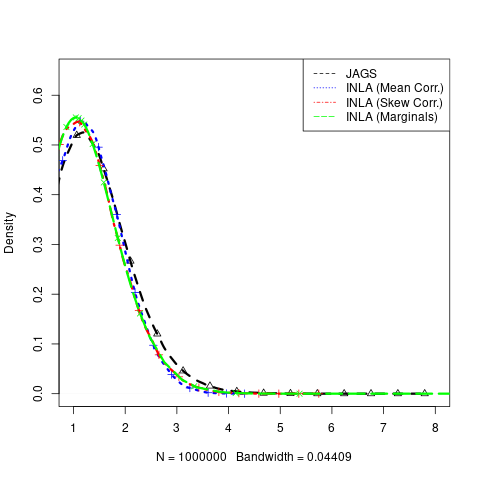}
\caption {A focus on the right tail of the linear predictor $\eta_{14}$ of the Binomial GLMM model where the skewness propagation to the tail is more evident. The skewness corrected marginal (red) from the SGC totally matches with the Simplified Laplace marginal result (green).}
\label{bslat14}       
\end{figure}

\begin{figure}[bp!]
  \centering
  \includegraphics[scale = 0.7]{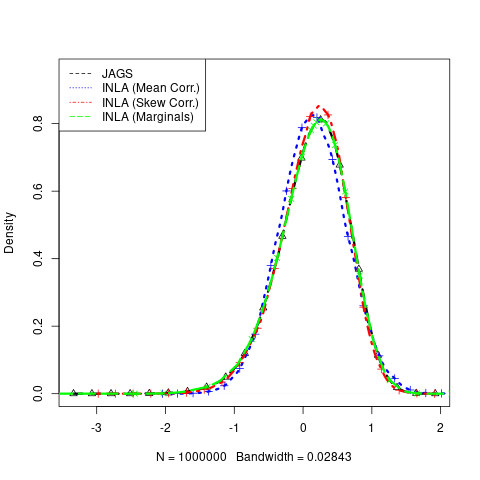}
\caption{Posterior marginal representation for linear predictor $\eta_9$ of the Poisson GLMM model with marginal skewness is around -0.4 for all the configuration points. The curves display the outcomes from different strategies: posterior marginal from JAGS (black), mean corrected (blue) and skewness corrected (red) marginal from the SGC and the Laplace posterior marginal (green) computed by INLA. }
\label{pla9}       
\end{figure}

\begin{figure}[bp!]
  \centering
  \includegraphics[scale = 0.7]{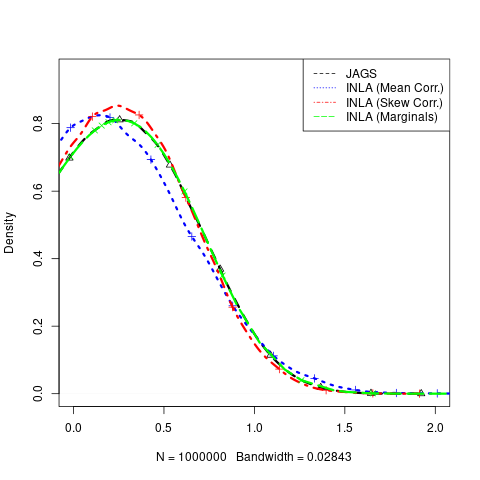}
\caption {A focus on the right tail of the linear predictor $\eta_9$ of the Poisson GLMM model where the skewness propagation to the tail is more evident. The skewness corrected marginal (red) from the SGC closely matches with the Laplace (green) and MCMC marginal (black) result.}
\label{plat9}       
\end{figure}

\begin{figure}[bp!]
  \centering
  \includegraphics[scale = 0.7]{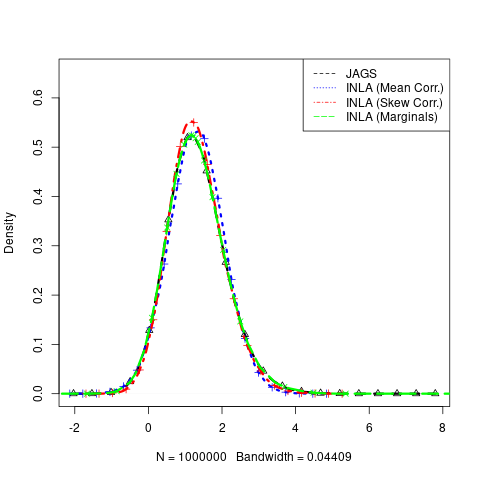}
\caption{Posterior marginal representation for linear predictor $\eta_{14}$ of the Binomial GLMM model with marginal skewness is around 0.33 for all the configuration points. The curves display the outcomes from different strategies: posterior marginal from JAGS (black), mean corrected (blue) and skewness corrected (red) marginal from the SGC and the Laplace posterior marginal (green) computed by INLA. }
\label{bla14}       
\end{figure}

\begin{figure}[bp!]
  \centering
  \includegraphics[scale = 0.7]{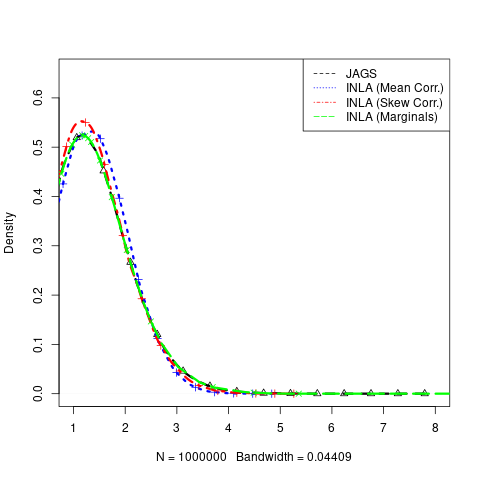}
\caption {A focus on the right tail of the linear predictor $\eta_{14}$ of the Binomial GLMM model where the skewness propagation to the tail is more evident. The skewness corrected marginal (red) from the SGC closely matches with the Laplace (green) and MCMC marginal (black) result.}
\label{blat14}       
\end{figure}

\noindent
The reader will note the substantial gain in speed with the INLA setup compared to JAGS: the joint INLA sampler is around 170 and 50 times faster for the mean and skewness corrected approach, respectively. To avoid being redundant, we show results of a single linear predictor marginal for both the Poisson and Binomial GLMM example. The marginal representation of these results is exactly extracted from the joint posterior outcome obtained from JAGS and R-INLA. \\ In particular, Figures \ref{psla9} and \ref{bsla14}  show the marginal results for the chosen linear predictor marginal using the Simplified Laplace strategy in INLA. This strategy is preferable since the results can be computed fast at the cost of some accuracy. However the skewness corrected marginals almost exactly match with the Simplified approximated marginals while the mean corrected results are more inaccurate. This difference is even more clear when we check the tail behaviour as shown in Figures \ref{pslat9} and \ref{bslat14}. More accurate results can be obtained through the Laplace strategy which require slightly more heavy computations. Figures \ref{pla9}, \ref{plat9}, \ref{bla14} and \ref{blat14} show the same comparative marginals results when fitting the model with the Laplace strategy in INLA. The joint posterior approximation does not retain the same accuracy as the approximated Laplace marginals by construction, but we still notice that the skewness corrected results lead to more appropriate conclusions opposed to the mean corrected ones.

\noindent
Positive (negative) marginal skewness leads to an underestimation (overestimation) of the real latent values as a result. This pattern is confirmed by noticing the accurate matching of the marginal MCMC distributions, the approximated INLA marginals, and the skewness corrected marginals from the INLA joint sampler. \\ In such extreme frameworks, the skewness effect on the marginals should not be ignored since the deviation can propagate to the joint distribution. \\ The results confirm that the Skew Gaussian Copula with its skewness correction applied on the transformed latent variables provide more accurate results when the marginal skewness is not negligible. The posterior marginal outcomes totally matches when the Simplified Laplace strategy is employed at the cost of some accuracy in terms of the sampling Monte Carlo Markov Chain derived truth. \\ On another note, increased precision for the skewness corrected marginals can be obtained by using the Laplace strategy. 

\subsection{Linear combination joint posterior inference}
\label{sec:3.2}

\begin{figure}[hbt!]
\centering
\includegraphics[scale = 0.8]{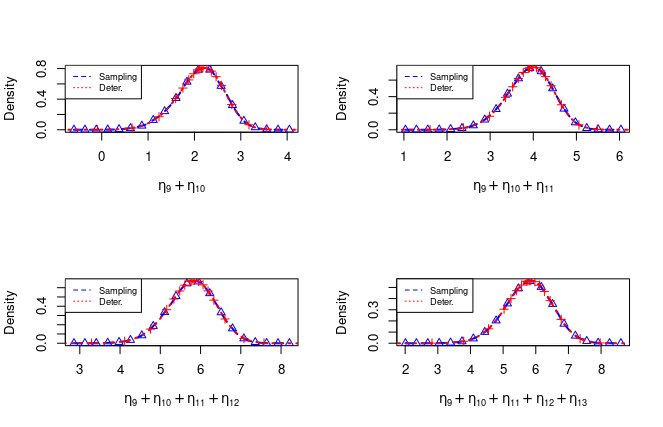}
\caption{One dimensional comparison for all the linear combinations obtained from the joint posterior using $10^5$ samples. Blue line marginal is obtained by sampling while red line represents the deterministic marginal result derived from the joint SGC class. All marginal linear combinations are skewed with marginal skewness evaluations $\gamma(\eta_9+\eta_{10} \vert \boldsymbol{y})=-0.33$, $\gamma(\eta_9+\eta_{10}+\eta_{11} \vert \boldsymbol{y})=-0.28$, $\gamma(\eta_9+\eta_{10}+\eta_{11}+\eta_{12} \vert \boldsymbol{y})=-0.21$ and $\gamma(\eta_9+\eta_{10}+\eta_{11}+\eta_{12}+\eta_{13} \vert \boldsymbol{y})=-0.18$. }
\label{dl1} 
\end{figure}

The use of the marginal skewness correction applied within the Skew Gaussian Copula class allows to get results more coherent to the marginal Simplified Laplace approximations computed by INLA. When the target of the posterior inferential analysis regards linear combinations of the latent field, we may also consider the alternative manipulation of the aforementioned class of joint densities pointed out in Section \ref{sec:2}. This method provides an alternative way to construct a surrogate Skew Gaussian Copula joint density for a set of pre-specified additive linear combinations where the marginals are modelled through Skew Normal densities (similar to the Simplified Laplace strategy). \\ User friendliness and speed are the key feature of this approach and an application showing the new R-INLA tools is exposed on Appendix A. \\ Similarly to the comparative analysis carried out for the joint posterior approximation, we consider again the Poisson hierarchical example of Section \ref{sec:4.1} and construct four increasing additive linear combinations in terms of the linear predictors $(\eta_9, \eta_{10}, \eta_{11}, \eta_{12}, \eta_{13})$. Our scope is to explore the posterior approximations of the considered linear combinations $\pi(\eta_9+\eta_{10} \vert \boldsymbol{y})$, $\pi(\eta_9+\eta_{10}+\eta_{11} \vert \boldsymbol{y})$, $\pi(\eta_9+\eta_{10}+\eta_{11}+\eta_{12} \vert \boldsymbol{y})$, $\pi(\eta_9+\eta_{10}+\eta_{11}+\eta_{12}+\eta_{13} \vert \boldsymbol{y})$ by using both the sample based joint posterior approximations and its manipulative deterministic version. Both approaches apply the marginal skewness correction and provide approximations of the target with the exception that the manipulative version does integrate out the hyperparameter uncertainty. This makes the joint posterior approximation by sampling more accurate and therefore closer to the truth. Figure \ref{dl1} directly shows a comparison between these two methods through the marginal representation of the joint posterior result for all linear combinations. The posterior marginal results are skewed by construction and both methods provide accurate matching results meaning that the manipulative joint approximations is a proper candidate for accurate linear combination inference. Hence the alternative approach can be a preferable choice since it avoids sampling. 

\noindent
Posterior summaries for the linear combinations of the Poisson example are shown in Table \ref{dt1} where both approahes are compared through Kullback Leibler distance with Monte Carlo sampling (\cite{kld2007}). This divergence measure represents a reasonable indicator to quantify how far the new manipulative tool is from its true sampling counterpart but must not be considered as an undeniable truth.  Since all the results come from not uni dimensional object which do not possess the usual marginal accuracy, it is worth to underline that the KLD measure has always a $10^{-3}$ order precision which means that the manipulative joint approximation does not lose much against the true sampling version. \\ This argument can be emphasized by a simple example by considering two generic well defined functions $f$ and $g$ that need to be compared. Then their KLD is defined as 

\begin{equation}
\text{KLD}(f \vert \vert g)= \int \pi_f \log \Bigl ( \frac{\pi_f}{\pi_g}\Bigr )= \text{E}_f \Bigl [\log \Bigl ( \frac{\pi_f}{\pi_g}\Bigr ) \Bigr ] 
\nonumber
\end{equation}

\noindent
Let us assume $\pi_f \sim \text{N}(0,1)$ and $\pi_g \sim \text{N}(\delta,1)$ with $\delta$ being a varying location constant. In this simple setting the resulting KLD is

\begin{equation}
\text{KLD}(f \vert \vert g)= \int \frac{1}{\sqrt{2\pi}} \exp \Bigl ( -\frac{x^2}{2}\Bigr ) \Bigl [-\frac{x^2}{2}+\frac{(x-\delta)^2}{2} \Bigr ]= \frac{\delta^2}{2}
\nonumber
\end{equation}

\noindent
Therefore a KLD equal to $10^{-3}$ would mean that the approximating function $\pi_g$ would have a location $\delta \approx 0.04$ which is really close to the truth. A $\delta \approx 0.01$ can be achieved with a KLD equal to $10^{-4}$. Similar results can be obtained if a varying standard deviation is assumed.

\begin{table}[bp!]
\centering
\caption{Posterior summaries and KLD evaluation for all one dimensional linear combinations in the Poisson hierarchical model.}
\label{dt2}       
\begin{tabular}{|l|l|l|l|l|l|l|l|}
\hline\noalign{\smallskip}
\textbf{Index} & \textbf{Mean} & \textbf{Sd} & \textbf{0.025quant} & \textbf{0.5quant} & \textbf{0.975quant} & \textbf{Mode} & \textbf{Kld} \\
\noalign{\smallskip}\hline\noalign{\smallskip}
$\eta_9+\eta_{10}$ & 2.116 & 0.505 &  1.042 & 2.146 &  3.025 & 2.209 & $1.26 \times 10^{-3}$ \\
$\eta_9+\eta_{10}+\eta_{11}$ & 3.912 & 0.537 &  2.783 & 3.939 &  4.893 & 3.995 & $1.18 \times 10^{-3}$ \\
$\sum_{i=9}^{12} \eta_i$ & 5.776 & 0.595 &  4.545 & 5.798 &  6.885 & 5.842 & $1.21 \times 10^{-3}$ \\
$\sum_{i=9}^{13} \eta_i$ & 5.772 & 0.718 & 4.300 & 5.794 &  7.120 & 5.838 & $1.25 \times 10^{-3}$ \\
\noalign{\smallskip}\hline
\end{tabular}
\label{dt1}
\end{table}

\noindent
Even the plot results confirms that the manipulative approximation through the SGC construction has a lack of accuracy that is practically negligible in this case with respect to the sampling counterpart. \\ As a conclusion, Table \ref{dspeed} report the computational time differences by using the joint sampler with a different number of samples and its joint manipulative alternative. From Table \ref{dspeed} we can observe a huge speed up of the manipulative version when applied to linear combinations compared to the sampling approach. This version is respectively 1100 and 4200 times faster on average than $10^3$ and $10^4$ samples from the joint posterior sampling based approach. This huge speed-up of the new tools is expected since we are only applying exact algebraic operations.

\begin{table}[bp!]
\centering
\caption{Speed Comparison between the joint deterministic algorithm and its sampling version using different sample sizes for computing all one dimensional linear combinations of the Poisson simulation. The performance results have been measured under 100 replications.}
\label{dspeed2}       
\begin{tabular}{|l|l|l|l|}
\hline\noalign{\smallskip}
\textbf{Method} & \textbf{Min} & \textbf{Mean} & \textbf{Max}  \\
\noalign{\smallskip}\hline\noalign{\smallskip}
\text{no samples} & 0.18ms & 0.36ms & 0.66ms \\
$10^3$ \text{samples} & 177ms & 396ms & 742ms \\
$10^4$ \text{samples} & 1056ms & 1524ms & 2024ms \\
\noalign{\smallskip}\hline
\end{tabular}
\label{dspeed}
\end{table}

\section{Discussion}

\begin{figure}[bp!]
\centering
\includegraphics[scale = 0.7]{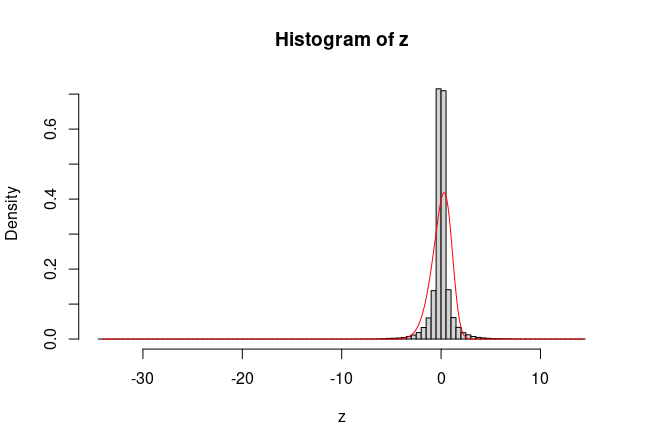}
\caption{Comparative example of a tri-product linear combination of random variables. The histogram shows the true result of the product $z=xyk$ while the red line represents the corresponding Skew Normal adaptation using the moments information. The true result is indeed far from the Skew Normal fit.}
\label{product}
\end{figure}

Various tools for joint posterior inference with INLA have been developed over time and here we add another one necessary for sampling from the skew-corrected joint posterior. The improved approximations from Section \ref{sec:1} quickly identify and solve deviations from Gaussianity by introducing skewness in the samples through the Gaussian copula. The skewness correction results in more accurate posterior results, even in reasonable extreme conditions, while keeping the procedure computationally efficient. \\ However, if the function is employed in a model fitting context where the grid is finer, that is, more configuration points for the hyperparameter set $\{\boldsymbol{\theta}_k, k=1,\dots,K\}$ are computed, then the additional computational burden becomes more evident. Onto this point, additional strategies may be needed such as distributing the computations amongst the configuration points within a parallel setting. \\ \\
If the model is well-defined and only additive linear combinations are involved then the manipulative joint version of the SGC as shown in Section \ref{sec:2} could be used to circumvent sampling while attaining approximate posterior joint inference. This alternative joint approach, in contrast to the sampling approach from Section \ref{sec:1}, is not hindered by a large set of configuration points for the hyperparameter set and produces inference almost instantly.\\
This provides an exciting avenue for future research into joint posterior inference for functionals of the latent field besides just linear combinations. The construction of an approximated Skew Gaussian Copula class through its moments evaluation for handling proper inference for linear combinations may be extended to products of the same but no clear shape of the resulting distributions is known so that the Skew Normal assumption is likely to fail. It can be shown that the product of more than one random variable largely deviate from Gaussianity and gets more and more spiky. Figure \ref{product} easily shows that even the Skew Normal density is not able to appropriately model the shape of a product. \\ Unless the distributional properties of the non-linear functional are known (near)exactly, the theoretical artifacts applied in Section \ref{sec:2} are of limited use. Even though products of random variables are important in several statistical contexts such as factor models or interaction terms, sums of random variables are more popular and widely used.

\noindent
We believe that these new tools for approximate (joint) Bayesian inference with INLA will enable the user to perform more accurate and ergonomic inference for various functionals from the joint posterior, while maintaining computational efficiency and ease of accessibility.

\begin{appendices}

\section{How to get joint approximations with R-INLA}
\label{AppA}

Let us briefly summarize what are the available INLA options that allow us to get joint posterior approximations for a Latent Gaussian Model:

\begin{description}
    \item[$\bullet$] a first joint approximation can be constructed through a sampling approach (as outlined in Section \ref{sec:3}) using the well established R-INLA function \texttt{inla.posterior.sample} which now also applies marginal skewness correction adjustments by using the option \texttt{skew.corr = TRUE}. \\ Many tutorials on this joint INLA posterior sampler can be found in \cite{krainski2018spatial} and \cite{martino2019integrated} and they accurately explain how to derive and interpret correct functionals of the results for prediction purposes.
    \item[$\bullet$] for additive linear combinations or functionals of a specific joint posterior subset of the latent field terms (as largely described in Section \ref{sec:2}), we can also derive deterministic approximations by constructing a similar joint posterior approximation with the same Gaussian copula structure but having Skew Normal marginals, that is, the mentioned \emph{Skew Gaussian Copula} (SGC) object.  This can be achieved by using the new available R-INLA functions \texttt{inla.tjmarginal} and \texttt{inla.1djmarginal}.
\end{description}

\noindent
Both approaches apply a simple and fast post-process of the main INLA output after fitting the model of interest. \\ \\ The R code below shows how to employ the new deterministic approach for approximating two additive linear combinations from a simple Poisson model application with three covariates.

\begin{verbatim}

# Poisson regression dataset (simulation) 

> nn = 50
> p = 3
> x1 <- rnorm(nn)
> x2 <- rnorm(nn)
> x3 <- rnorm(nn)
> eta = 1+x1+x2+x3
> y = rpois(nn, lambda = exp(eta)) 

> data = data.frame(y = y, x1 = x1, x2 = x2, x3 = x3)

> sel = list('(Intercept)'= 1, x1 = 1, x2 = 1, x3 = 1)

# Fitting the data into an INLA framework

> mod.ex <- inla(y ~ 1+x1+x2+x3, 
>                family = "poisson", 
>                data = data, 
>                selection = sel,
>                control.compute = list(config = TRUE))

# Getting deterministic approximations from the SNTGC object

> Lin.idx = matrix(c(0, 1, 1, 0, 0, 1, 1, 1), nrow = 2, ncol = p+1,  byrow = T)

> Lin.idx
     [,1] [,2] [,3] [,4]
[1,]    0    1    1    0
[2,]    0    1    1    1

> mod.det = inla.tjmarginal(jmarginal = mod.ex$selection, A = Lin.idx)

$names
[1] "Lin:1" "Lin:2"

$mean
         [,1]
[1,] 1.971917
[2,] 3.013960

$cov.matrix
            [,1]        [,2]
[1,] 0.002613337 0.002072906
[2,] 0.002072906 0.004420322

$skewness
[1] 0.06482934 0.03272285

> dm = inla.1djmarginal(jmarginal = mod.det)

# Plot the deterministic results

> plot(dm$`Lin:1`, type = 'l')
> plot(dm$`Lin:2`, type = 'l')

\end{verbatim}

\noindent
Through the \texttt{selection} list option within the main INLA function call, we are able to extract all the moments and correlation structure of the parameters of interest: in this case given by the three model covariates $x_1$, $x_2$ and $x_3$. \\ As an example, we focus on getting posterior approximations for the two specified linear combinations $x_1+x_2$ and $x_1+x_2+x_3$ from the joint posterior structure of the model. First we use the new function \texttt{inla.tjmarginal} to construct the SGC object structure with its moments and correlation matrix amongst the two linear combinations whose indexes are specified into \texttt{Lin.idx}. Then we use the last new function \texttt{inla.1djmarginal} to derive the Skew Normal approximations of the involved linear combinations so that we can achieve deterministic approximations for $\pi(x_1+x_2 \vert \boldsymbol{y})$ and $\pi(x_1+x_2+x_3 \vert \boldsymbol{y})$. \\ The results can be easily compared to the ones from the usual joint sampler \texttt{inla.posterior.sample} function. \\ More examples, applications of these new functions can be found on the tutorial vignette available on the R-INLA official website.

\section{A look at the computational advantages}
\label{AppB}

One may wonder why we should use the two-way strategy summarised in Proposition \ref{map2d} if we already know the solution of the problem stated in~\eqref{quantile:eq}. The answer entirely relies on the computational gain which allows a feasible and reasonable application. Our strategy has been compared to the most used default available options in R language and the achieved results are satisfying for our practical needs. \\ \\
The constants in the first step of the strategy determine many important numerical details to ensure a good balance of accuracy and speed for the next steps. As already remarked, few fixed global constants are constructed with the aim to control the range, the number of evaluations, and digits accuracy for the interpolant functions. The default numbers can be adjusted within the original code, but any small change can affect both accuracy and speed performances greatly. It is strongly suggested not to mess with these values since the default is enough for general purposes. \\ For example, by default, the built-in functions construct the interpolants by computing all possible skewness evaluations with a precision up to the second digit. Increasing the number of digits leads to an increment in the number of interpolant functions to be calculated. Although this enlarges the marginal fit accuracy of a negligible order, the computational burden becomes much heavier. \\ By keeping these assumptions on the global constants, the initial problem is solved by tabulating all possible solutions from the standard quantile function \texttt{qsn()} in R and then build and save all the interpolants in the cache. Worth to underline that this part is only computed once and continues to exist in the local R global environment session. \\ Any marginal skewness adjustment is then applied to the samples by calling the respective correct interpolants stored in the cache. As a matter of fact, this new whole computational approach turns out to be both automatic and fast. \\ \\ As it appears in Table \ref{tab:speed}, the standard available function \texttt{qsn()} and the new faster strategy as described in Proposition \ref{map2d}, are compared under 100 replications each and the respective results for the pdf, cdf and quantile equation of the Skew Normal distribution are reported. In Table \ref{tab:speed}, $\tilde{F}^{-1}$ refers to $\tilde{F}^{-1}[\Phi(\boldsymbol{x})]$. \\ Note that even the pdf and cdf functions are computed using interpolants, since the available functions cannot evaluate many different Skew Normal densities at the same time. \\ Based on $N = 10^6$ evaluations of the latent field points, it is evident that the biggest gain in speed is achieved by the alternative cdf and quantile versions. Indeed, the speed-up coming from the new quantile version is approximately 15 times faster than \texttt{qsn()} on average. \\ This is enough evidence to state that the two-way strategy grants better computational advantages than available R packages and should be preferred in this case.

\begin{table}[h]
\centering
\caption{Speed Time Results comparison between the standard \texttt{qsn()} function and the new faster strategy. Function evaluations based on 100 replications and $N = 10^6$ points. Function $\boldsymbol{\tilde{f}_{\text{std}}}$, $\boldsymbol{\tilde{F}_{\text{std}}}$ and $\boldsymbol{\tilde{F}^{-1}_{\text{std}}}$ are respectively the pdf, cdf and quantile function of the Skew Normal distribution implemented in \texttt{qsn()}. Accordingly $\boldsymbol{\tilde{f}_{\text{fast}}}$, $\boldsymbol{\tilde{F}_{\text{fast}}}$ and $\boldsymbol{\tilde{F}^{-1}_{\text{fast}}}$ relate to the alternative faster strategy.}
\label{tab:speed} 
\begin{tabular}{|l|l|l|l|}
\hline\noalign{\smallskip}
\textbf{} & \textbf{Min} & \textbf{Mean} & \textbf{Max}  \\
\noalign{\smallskip}\hline\noalign{\smallskip}
  $\boldsymbol{\tilde{f}_{\text{std}}}$ & 1.51ms & 2.23ms & 7.32ms \\
$\boldsymbol{\tilde{f}_{\text{fast}}}$ & 1.27ms & 2.1ms & 6.1ms \\
$\boldsymbol{\tilde{F}_{\text{std}}}$ & 6221$\mu s$ & 9144$\mu s$ &  19973$\mu s$ \\
$\boldsymbol{\tilde{F}_{\text{fast}}}$ & 817$\mu s$ & 1243$\mu s$ & 4143$\mu s$ \\
$\boldsymbol{\tilde{F}^{-1}_{\text{std}}}$ & 22.94s & 25.27s & 30.46s \\
$\boldsymbol{\tilde{F}^{-1}_{\text{fast}}}$ & 1.61s & 1.78s & 3.18s \\
\noalign{\smallskip}\hline
\end{tabular}
\end{table}

\end{appendices}

\bibliographystyle{apalike}
\bibliography{References.bib}   

\begin{thebibliography}{}

\bibitem[Alvaro-Meca et~al., 2013]{alvaro2013}
Alvaro-Meca, A., Akerkar, R., Alvarez-Bartolome, M., Gil-Prieto, R., Rue, H.,
  and de~Miguel, A.~G. (2013).
\newblock Factors involved in health related transitions after curative
  resection for pancreatic cancer 10 year experience: A multi state model.
\newblock {\em Cancer Epidemiology}, 37(1):91--96.

\bibitem[Azzalini and Capitanio, 1999]{Azzalini_1999}
Azzalini, A. and Capitanio, A. (1999).
\newblock Statistical applications of the multivariate skew normal
  distribution.
\newblock {\em Journal of the Royal Statistical Society: Series B (Statistical
  Methodology)}, 61(3):579–602.

\bibitem[Bauer et~al., 2016]{bauer2016}
Bauer, C., Wakefield, J., Rue, H., Self, S., Feng, Z., and Wang, Y. (2016).
\newblock Bayesian penalized spline models for the analysis of spatio-temporal
  count data.
\newblock {\em Statistics in Medicine}, 35(11):1848--1865.

\bibitem[Beguin et~al., 2012]{beguin2012}
Beguin, J., Martino, S., and Rue, H. (2012).
\newblock Hierarchical analysis of spatially autocorrelated ecological data
  using integrated nested {L}aplace approximation.
\newblock {\em Methods in Ecology and Evolution}, 3(5):921--929.

\bibitem[Dawkins et~al., 2019]{dawkins2019}
Dawkins, L., Williamson, D., Mengersen, K., and Shaddick, G. (2019).
\newblock 'where is the clean air?' a bayesian decision framework for
  personalised route selection using inla.

\bibitem[Ferkingstad et~al., 2017]{ferkingstad2017}
Ferkingstad, E., Held, L., and Rue, H. (2017).
\newblock Fast and accurate {B}ayesian model criticism and conflict diagnostics
  using {R-INLA}.
\newblock {\em Stat}, 6(1):331--344.

\bibitem[Ferkingstad and Rue, 2015]{ferkingstad2015latent}
Ferkingstad, E. and Rue, H. (2015).
\newblock Improving the {INLA} approach for approximate {B}ayesian inference
  for latent {G}aussian models.
\newblock {\em Electronic Journal of Statistics}, 9:2706--2731.

\bibitem[Ghorbanzadeh et~al., 2017]{ghorbanzadeh_2017}
Ghorbanzadeh, D., Durand, P., and Jaupi, L. (2017).
\newblock Generating the skew normal random variable.

\bibitem[Gomez-Rubio et~al., 2021]{gomezrubio2021}
Gomez-Rubio, V., Bivand, R.~S., and Rue, H. (2021).
\newblock Estimating spatial econometrics models with integrated nested laplace
  approximation.

\bibitem[Gómez-Rubio and Palmí-Perales, 2019]{gomez2019}
Gómez-Rubio, V. and Palmí-Perales, F. (2019).
\newblock Multivariate posterior inference for spatial models with the
  integrated nested laplace approximation.
\newblock {\em Journal of the Royal Statistical Society: Series C (Applied
  Statistics)}, 68(1):199--215.

\bibitem[{Hershey} and {Olsen}, 2007]{kld2007}
{Hershey}, J.~R. and {Olsen}, P.~A. (2007).
\newblock Approximating the kullback leibler divergence between gaussian
  mixture models.
\newblock In {\em 2007 IEEE International Conference on Acoustics, Speech and
  Signal Processing - ICASSP '07}, volume~4, pages IV--317--IV--320.

\bibitem[Holand et~al., 2013]{holand2013}
Holand, A.~M., Steinsland, I., Martino, S., and Jensen, H. (2013).
\newblock Animal models and integrated nested {L}aplace approximations.
\newblock {\em {G3: Genes|Genomics|Genetics}}, 3(8):1241--1251.

\bibitem[Huang et~al., 2017]{huang2017}
Huang, J., Malone, B., Minasny, B., Mcbratney, A., and Triantafilis, J. (2017).
\newblock Evaluating a bayesian modelling approach (inla-spde) for
  environmental mapping.
\newblock {\em Science of The Total Environment}, 609:621--632.

\bibitem[Illian et~al., 2012]{illian2012}
Illian, J.~B., S{\o}rbye, S.~H., Rue, H., and Hendrichsen, D.~K. (2012).
\newblock Using {INLA} to fit a complex point process model with temporally
  varying effects -- a case study.
\newblock {\em Journal of Environmental Statistics}, 3(7).

\bibitem[Krainski et~al., 2018]{krainski2018spatial}
Krainski, E.~T., G{\'o}mez-Rubio, V., Bakka, H., Lenzi, A., Castro-Camilio, D.,
  Simpson, D., Lindgren, F., and Rue, H. (2018).
\newblock {\em Advanced Spatial Modeling with Stochastic Partial Differential
  Equations using {R} and {INLA}}.
\newblock CRC press.
\newblock Github version www.r-inla.org/spde-book.

\bibitem[Li et~al., 2012]{yli2012}
Li, Y., Brown, P., Rue, H., {al-Maini}, M., and Fortin, P. (2012).
\newblock Spatial modelling of {Lupus} incidence over 40 years with changes in
  census areas.
\newblock 61:99--115.

\bibitem[Martino et~al., 2010a]{martino2010est}
Martino, S., Aas, K., Lindqvist, O., Neef, L.~R., and Rue, H. (2010a).
\newblock Estimating stochastic volatility models using integrated nested
  {L}aplace approximation.
\newblock {\em The European Journal of Finance}, pages 1--17.

\bibitem[Martino et~al., 2010b]{martino2010approx}
Martino, S., Akerkar, R., and Rue, H. (2010b).
\newblock Approximate {B}ayesian inference for survival models.
\newblock 28(3):514--528.

\bibitem[Martino and Riebler, 2019]{martino2019integrated}
Martino, S. and Riebler, A. (2019).
\newblock Integrated nested laplace approximations (inla).

\bibitem[Martins et~al., 2013]{martins2013new}
Martins, T.~G., Simpson, D., Lindgren, F., and Rue, H. (2013).
\newblock Bayesian computing with {INLA}: {N}ew features.
\newblock 67:68--83.

\bibitem[Meehan et~al., 2019]{meehan2019}
Meehan, T.~D., Michel, N.~L., and Rue, H. (2019).
\newblock Spatial modeling of audubon christmas bird counts reveals fine-scale
  patterns and drivers of relative abundance trends.
\newblock {\em Ecosphere}, 10(4).
\newblock Article e02707.

\bibitem[Muff et~al., 2015]{muff2015}
Muff, S., Riebler, A., Rue, H., Saner, P., and Held, L. (2015).
\newblock Bayesian analysis of measurement error models using integrated nested
  {L}aplace approximations.
\newblock 64(2):231--252.

\bibitem[Nelsen, 1999]{nelsen_1999_an}
Nelsen, R.~B. (1999).
\newblock {\em An introduction to copulas}.
\newblock Springer.

\bibitem[Peluso et~al., 2020]{peluso2020}
Peluso, S., Mira, A., Rue, H., Tierney, N.~J., Benvenuti, C., Cianella, R.,
  Caputo, M.~L., and Auricchio, A. (2020).
\newblock A bayesian spatiotemporal statistical analysis of out-of-hospital
  cardiac arrests.
\newblock {\em Biometrical Journal}, 62(4):1105--1119.

\bibitem[Pereira et~al., 2017]{pereira2017}
Pereira, S., Turkman, K.~F., Correia, L., and Rue, H. (2017).
\newblock Unemployment estimation: Spatial point referenced methods and models.

\bibitem[Phillips, 2010]{sym2010}
Phillips, K. (2010).
\newblock R functions to symbolically compute the central moments of the
  multivariate normal distribution.
\newblock {\em Journal of Statistical Software}, 33.

\bibitem[Plummer, 2003]{plummer_2003_dsc}
Plummer, M. (2003).
\newblock Dsc 2003 working papers jags: A program for analysis of bayesian
  graphical models using gibbs sampling.

\bibitem[Quiroz et~al., 2015]{quiroz2015}
Quiroz, Z., Prates, M.~O., and Rue, H. (2015).
\newblock A {B}ayesian approach to estimate the biomass of anchovies in the
  coast of {Per\'u}.
\newblock 71(1):208--217.

\bibitem[Riebler et~al., 2012a]{riebler2012est}
Riebler, A., Held, L., and Rue, H. (2012a).
\newblock Estimation and extrapolation of time trends in registry data -
  {B}orrowing strength from related populations.
\newblock 6(1):304--333.

\bibitem[Riebler et~al., 2012b]{riebler2012gender}
Riebler, A., Held, L., Rue, H., and Bopp, M. (2012b).
\newblock Gender-specific differences and the impact of family integration on
  time trends in age-stratified swiss suicide rates.
\newblock 175(2):473--490.

\bibitem[Rue and Held, 2005]{rue2005gmrf}
Rue, H. and Held, L. (2005).
\newblock {\em Gaussian {M}arkov Random Fields: {T}heory and Applications},
  volume 104 of {\em Monographs on Statistics and Applied Probability}.
\newblock Chapman \& Hall, London.

\bibitem[Rue and Martino, 2007]{martino2007}
Rue, H. and Martino, S. (2007).
\newblock Approximate bayesian inference for hierarchical gaussian markov
  random fields.
\newblock {\em Journal of Statistical Planning and Inference}, 137:3177--3192.

\bibitem[Rue et~al., 2009]{rue2009inla}
Rue, H., Martino, S., and Chopin, N. (2009).
\newblock Approximate {B}ayesian inference for latent {G}aussian models using
  integrated nested {L}aplace approximations (with discussion).
\newblock 71(2):319--392.

\bibitem[Rue et~al., 2017]{rue2017computing}
Rue, H., Riebler, A., S{\o}rbye, S.~H., Illian, J.~B., Simpson, D.~P., and
  Lindgren, F.~K. (2017).
\newblock Bayesian computing with {INLA}: {A} review.
\newblock {\em Annual Reviews of Statistics and Its Applications},
  4(March):395--421.

\bibitem[Ruiz-C\'ardenas et~al., 2012]{rruiz2012}
Ruiz-C\'ardenas, R., Krainski, E.~T., and Rue, H. (2012).
\newblock Direct fitting of dynamic models using integrated nested {L}aplace
  approximations - {INLA}.
\newblock 56(6):1808--1828.

\bibitem[Rustand et~al., 2021]{rustand2021bayesian}
Rustand, D., van Niekerk, J., Rue, H., Tournigand, C., Rondeau, V., and
  Briollais, L. (2021).
\newblock Bayesian estimation of two-part joint models for a longitudinal
  semicontinuous biomarker and a terminal event with r-inla: Interests for
  cancer clinical trial evaluation.

\bibitem[Seppä et~al., 2018]{seppa2018}
Seppä, K., Rue, H., Hakulinen, T., Läärä, E., Sillanpää, M.~J., and
  Pitkäniemi, J. (2018).
\newblock Estimating multilevel regional variation in excess mortality of
  cancer patients using integrated nested laplace approximation.
\newblock {\em Statistics in Medicine}, 38:778--791.

\bibitem[S{\o}rbye et~al., 2019]{sorbye2019}
S{\o}rbye, S.~H., Myrvoll-Nilsen, E., and Rue, H. (2019).
\newblock An approximate fractional {G}aussian noise model with {$O(n)$}
  computational cost.
\newblock {\em Statistics and Computing}, 29(4):821--833.

\bibitem[van Niekerk et~al., 2019]{vanniekerk2019new}
van Niekerk, J., Bakka, H., Rue, H., and Schenk, O. (2019).
\newblock New frontiers in bayesian modeling using the inla package in r.

\bibitem[Wakefield et~al., 2016]{wakefield_2016_spatial}
Wakefield, J., Simpson, D., and Godwin, J. (2016).
\newblock Spatial modeling, with application to complex survey data: Discussion
  of "model-based geostatistics for prevalence mapping in low-resource
  settings", by diggle and giorgi.

\bibitem[Yuan et~al., 2017]{yyuan2017}
Yuan, Y., Bachl, F.~E., Borchers, D.~L., Lindgren, F., Illian, J.~B., Buckland,
  S.~T., Rue, H., and Gerrodette, T. (2017).
\newblock Point process models for spatio-temporal distance sampling data from
  a large-scale survey of blue whales.
\newblock {\em Annals of Applied Statistics}, 11(4):2270--2297.

\bibitem[Yue et~al., 2019]{yyue2019}
Yue, Y.~R., Bolin, D., Rue, H., and Wang, X. (2019).
\newblock Bayesian generalized two-way {ANOVA} modeling for functional data
  using {INLA}.
\newblock {\em Statistica Sinica}, 29(2):741--767.

\end{thebibliography}

\end{document}